\documentclass[aps, prd, 10pt,  amsmath, amssymb,  eqsecnum,  nofootinbib]{revtex4-1}

\usepackage[T1]{fontenc}
\usepackage{bm,mathtools}
\usepackage{xcolor}
\usepackage{graphicx}
\providecommand{\ii}{\mathrm{i}}
\newlength{\FRDiagramColumn}
\newbox\FigurePDFSeedBox
\makeatletter
\newcommand{\ContinuedFigureCaption}[1]{%
  \@makecaption{\fnum@figure}{#1}%
}
\define@key{ManuscriptFigurePDF}{scale}{\def\FigurePDFScale{#1}}
\newcommand{\ManuscriptFigurePDF}[2][]{%
  \par
  \begingroup
  \setkeys{ManuscriptFigurePDF}{scale=1,#1}
  \csname FigurePDF@#2\endcsname
  \setbox\z@=\hbox{%
    \kern\FigurePDFX sp\relax
    \raise\FigurePDFY sp\hbox{\includegraphics{figures/#2.pdf}}%
  }%
  \wd\z@=\linewidth
  \ht\z@=\FigurePDFHeight sp\relax
  \dp\z@=0pt
  \ifdim\FigurePDFScale pt=1pt
  \else
    \setbox\z@=\hbox to\linewidth{\hss\scalebox{\FigurePDFScale}{\box\z@}\hss}
  \fi
  \nointerlineskip\box\z@
  \ifnum\pdfstrcmp{#2}{figure_10_four_point_vertices_part1}=0
    \setbox\FigurePDFSeedBox=\hbox{\pdfliteral direct{3 Tr}\rlap{\kern 12607032sp\raise 894313sp\hbox{\fontsize{9}{10}\selectfont$\displaystyle16$}}\kern 14577572sp\raise 1294345sp\hbox{\fontsize{9}{10}\selectfont$\displaystyle \Bigl[(c_s^2+f_3)\left(k_1^{a_2}k_2^{a_1}+\delta^{a_1a_2}k_1^0k_2^0\right)-(2c_s^2+4f_3+f_4)k_1^{a_1}k_2^{a_2}\Bigr]\Bigg\}$}\pdfliteral direct{0 Tr}}%
    \ifdim\FigurePDFScale pt=1pt
    \else
      \wd\FigurePDFSeedBox=\linewidth
      \setbox\FigurePDFSeedBox=\hbox to\linewidth{\hss\scalebox{\FigurePDFScale}{\box\FigurePDFSeedBox}\hss}
    \fi
    \wd\FigurePDFSeedBox=0pt\ht\FigurePDFSeedBox=0pt\dp\FigurePDFSeedBox=0pt
    \nointerlineskip\box\FigurePDFSeedBox
  \fi
  \dimen@=\FigurePDFDepth sp\relax
  \dimen@=\FigurePDFScale\dimen@
  \edef\FigurePDFDepth{\number\dimen@}
  \expandafter\endgroup\expandafter\prevdepth\FigurePDFDepth sp\relax
}
\makeatother
\expandafter\def\csname FigurePDF@figure_01_scales\endcsname{\def\FigurePDFX{2499707}\def\FigurePDFY{92726}\def\FigurePDFHeight{5351346}\def\FigurePDFDepth{0}}
\expandafter\def\csname FigurePDF@figure_02_propagator_rules\endcsname{\def\FigurePDFX{5196759}\def\FigurePDFY{722684}\def\FigurePDFHeight{9006270}\def\FigurePDFDepth{679322}}
\expandafter\def\csname FigurePDF@figure_03_three_point_vertices\endcsname{\def\FigurePDFX{526254}\def\FigurePDFY{882105}\def\FigurePDFHeight{24632315}\def\FigurePDFDepth{1850698}}
\expandafter\def\csname FigurePDF@figure_04_exact_multiplet_propagator\endcsname{\def\FigurePDFX{2105016}\def\FigurePDFY{277995}\def\FigurePDFHeight{2378096}\def\FigurePDFDepth{1041592}}
\expandafter\def\csname FigurePDF@figure_05_tt_external_conversion\endcsname{\def\FigurePDFX{11446026}\def\FigurePDFY{82604}\def\FigurePDFHeight{6524301}\def\FigurePDFDepth{0}}
\expandafter\def\csname FigurePDF@table_II_phonon_emission\endcsname{\def\FigurePDFX{460472}\def\FigurePDFY{457854}\def\FigurePDFHeight{32159747}\def\FigurePDFDepth{1947594}}
\expandafter\def\csname FigurePDF@figure_06_general_ward_identity\endcsname{\def\FigurePDFX{7170212}\def\FigurePDFY{186736}\def\FigurePDFHeight{3997163}\def\FigurePDFDepth{0}}
\expandafter\def\csname FigurePDF@figure_07_three_point_multiplet_ward\endcsname{\def\FigurePDFX{7893811}\def\FigurePDFY{310888}\def\FigurePDFHeight{4713351}\def\FigurePDFDepth{0}}
\expandafter\def\csname FigurePDF@table_III_three_point_currents\endcsname{\def\FigurePDFX{263127}\def\FigurePDFY{452573}\def\FigurePDFHeight{31496648}\def\FigurePDFDepth{2415023}}
\expandafter\def\csname FigurePDF@figure_08_three_point_diagonal_ward\endcsname{\def\FigurePDFX{4736287}\def\FigurePDFY{-782782}\def\FigurePDFHeight{11645055}\def\FigurePDFDepth{5675071}}
\expandafter\def\csname FigurePDF@figure_09_kmoc_cut\endcsname{\def\FigurePDFX{12301189}\def\FigurePDFY{28005}\def\FigurePDFHeight{5088285}\def\FigurePDFDepth{0}}
\expandafter\def\csname FigurePDF@figure_10_four_point_vertices_part1\endcsname{\def\FigurePDFX{855163}\def\FigurePDFY{467286}\def\FigurePDFHeight{37826410}\def\FigurePDFDepth{7763271}}
\expandafter\def\csname FigurePDF@figure_10_four_point_vertices_part2\endcsname{\def\FigurePDFX{263127}\def\FigurePDFY{734590}\def\FigurePDFHeight{35922916}\def\FigurePDFDepth{2160760}}
\expandafter\def\csname FigurePDF@figure_11_four_point_multiplet_ward\endcsname{\def\FigurePDFX{460472}\def\FigurePDFY{-86531}\def\FigurePDFHeight{7933928}\def\FigurePDFDepth{3819508}}
\expandafter\def\csname FigurePDF@figure_12_four_point_diagonal_ward\endcsname{\def\FigurePDFX{1710326}\def\FigurePDFY{-107783}\def\FigurePDFHeight{33962253}\def\FigurePDFDepth{16833670}}

\usepackage[
  colorlinks=true,
  linkcolor=black,
  citecolor=blue,
  urlcolor=black
]{hyperref}

\allowdisplaybreaks[3]

\newcommand{\draftnoteMS}[1]{%
  \ifmmode
    \text{\textcolor{magenta}{\textit{MS: #1}}}%
  \else
    \textcolor{magenta}{\textit{MS: #1}}%
  \fi
}

\newcommand{\draftnote}[1]{%
  \ifmmode
    \text{\textcolor{blue}{\textit{#1}}}%
  \else
    \textcolor{blue}{\textit{#1}}%
  \fi
}

\newcommand{\app}[1]{Appendix~\ref{#1}}
\newcommand{\eqn}[1]{Eq.~\eqref{#1}}

\newcommand{\pol}{\varepsilon}

\begin{document}

\title{Environmental Effects in Post-Minkowskian Dynamics: Effective Field Theory, Feynman Rules, and Ward Identities for Compact Objects in Relativistic Fluids}

\author{Zvi Bern}
\author{Samuel Degen}
\author{Enrico Herrmann}
\author{Mikhail P. Solon}
\author{Anna M. Wolz}

\affiliation{
Mani L. Bhaumik Institute for Theoretical Physics,
University of California at Los Angeles,
Los Angeles, CA 90095, USA}

\date{\today}

\begin{abstract}
\vskip 2 cm 
\centerline{\bf Abstract}
We develop a hydrodynamic effective field theory for the gravitational dynamics of compact objects moving through an inviscid fluid environment. The fluid is described by the effective theory of perfect fluids, whose Goldstone phonons are kept as explicit fields coupled to gravity. We establish a consistent power counting for the resulting Feynman rules and derive the propagators and interaction vertices through four points in $D$ dimensions, together with the generalized on-shell Ward identities they obey. These identities relate amplitudes with an external graviton to those with the graviton replaced by a phonon, and thus provide nontrivial checks on higher-order calculations. As an application, we recover the leading-order relativistic dynamical-friction force from the tree-level amplitude for single-phonon emission. This work is a first step towards a toolkit for incorporating environmental effects into the scattering-amplitude pipeline used in post-Minkowskian calculations.

\end{abstract}

\maketitle

\tableofcontents
\newpage


\section{Introduction}

The direct detection of gravitational waves~\cite{LIGOScientific:2016aoc, LIGOScientific:2017vwq} and the prospect of rapidly improving observational precision~\cite{Punturo:2010zz, LISA:2024hlh, Reitze:2019iox, LIGOasharp, Abac:2025saz} call for commensurate advances in theoretical predictions of gravitational-wave signals. Most waveform calculations assume that binary inspirals occur in vacuum, an approximation well justified for the signals accessible to current detectors.  However, the greater sensitivity of next-generation space-based observatories and their ability to track inspirals over extended periods make it important to determine whether even small environmental effects can accumulate sufficiently to affect waveform modeling and signal interpretation (see, e.g., Refs.~\cite{Barausse:2014tra, Vicente:2019ilr, Toubiana:2020drf, Cole:2022yzw, Zwick:2022dih, LISAConsortiumWaveformWorkingGroup:2023arg, Chen:2025qyj, Garg:2024qxq, LISA:2022yao, Yunes:2011ws, Cardoso:2019rou}). 

Environmental effects such as accretion, dynamical friction, and planetary migration introduce dissipative forces into binary inspirals, with potentially important consequences for long-lived systems such as extreme-mass-ratio inspirals (EMRIs)~\cite{Petrich:1989, Kocsis:2011dr, Barausse:2014tra, Dotti:2005kq, Caputo:2020irr, Khalvati:2024tzz, Derdzinski:2020wlw}. However, calculating these effects on binary inspiral dynamics to higher orders in Newton's constant $G$ remains an open problem. In this work, we take initial steps toward developing an effective field theory (EFT) to compute long-range corrections to the gravitational dynamics of compact objects immersed in an inviscid fluid environment. The fluid is described by the effective theory of perfect fluids~\cite{Dubovsky:2005xd, Endlich:2010hf, Dubovsky:2011sj} (see Refs.~\cite{Herglotz:1911, Taub:1954zz, Schutz:1970my, Soper:1976bb} for early work), with the phonons kept as explicit fields coupled to gravity, and the compact object by a massive scalar field, as in the amplitudes approach to post-Minkowskian dynamics. Our ultimate goal is to place environmental effects under the same perturbative control as vacuum scattering, so that they can be incorporated into the scattering-amplitude pipeline used in higher-order post-Minkowskian calculations. In the vacuum problem, the interplay among several formulations has been highly productive: worldline EFT~\cite{Goldberger:2004jt, Dlapa:2021vgp}, worldline quantum field theory~\cite{Jakobsen:2023ndj, Driesse:2026qiz}, and amplitude-based approaches~\cite{Cheung:2018wkq, Kosower:2018adc, Bern:2019nnu, Damgaard:2023ttc}. We hope the present formulation encourages parallel treatments of environmental effects within each of these frameworks, as is already underway on the worldline side~\cite{Modrekiladze:2026twz, Modrekiladze:2026drr}. 

Our formulation is designed to facilitate the extension to higher orders and more general settings by drawing on field-theoretic methods developed for gravitational dynamics; see Ref.~\cite{Buonanno:2022pgc} for a review. These methods fit naturally within EFT and scattering amplitudes~\cite{Iwasaki:1971vb, Hiida:1972xs, Goldberger:2004jt, Neill:2013wsa, Bjerrum-Bohr:2013bxa, Cheung:2018wkq, Kosower:2018adc, Bern:2019nnu}, and include gauge-invariant on-shell amplitudes, generalized unitarity for constructing loop integrands from tree amplitudes~\cite{Bern:1994zx, Bern:1994cg}, double-copy relations between gauge and gravity theories~\cite{Kawai:1985xq, Bern:2008qj, Bern:2010ue, Bern:2019prr}, and multiloop integration techniques~\cite{Chetyrkin:1981qh, Laporta:2000dsw, Kotikov:1990kg, Bern:1993kr, Remiddi:1997ny, Henn:2013pwa}. The application of these particle-physics tools has driven rapid progress in post-Minkowskian gravitational dynamics~\cite{Bern:2019nnu, Bern:2019crd, Bern:2021dqo, Bern:2021yeh, Dlapa:2021vgp, Dlapa:2021npj, Manohar:2022dea, Jakobsen:2023ndj, Damgaard:2023ttc}, recently reaching the fifth post-Minkowskian order for nonspinning bodies~\cite{Driesse:2024xad, Bern:2024adl, Driesse:2024feo, Bern:2025wyd, Bern:2025zno, Driesse:2026qiz}. Related amplitude-based approaches have also been applied to three-body dynamics, spin, and finite-size tidal effects; see, for example, Refs.~\cite{Buonanno:2022pgc, Travaglini:2022uwo, Correia:2024yfx}.  The present work places fluid-induced environmental effects within a framework compatible with this same collection of tools.

One such long-range environmental effect useful for illustrating computations with this EFT is dynamical friction, the drag force on a compact object moving in a medium. Dynamical friction results from a massive object gravitationally perturbing the environment and generating a trailing overdense wake, whose gravitational attraction pulls back on the object, transferring energy and momentum from the orbit to the environment. This is one of the oldest studied environmental effects, first examined by Chandrasekhar analytically for a nonrelativistic collisionless system~\cite{Chandrasekhar:1943ys}, and subsequently extended to include post-Newtonian corrections~\cite{Lee:1969} and motion through gaseous and relativistic fluid media~\cite{Ostriker:1998fa, Kim:2007zb, Katz:2019rgf, Barausse:2007ph, desjacques_analytic_2022, Correia:2022gcs, Chiari:2022kas, Cardoso:2022whc}. More recently, growing interest in environmental effects on extreme-mass-ratio inspirals has further motivated studies of dynamical
friction for a body orbiting a primary black hole~\cite{Brito:2023pyl, Dyson:2025dlj, Datta:2025ruh, Datta:2026eqj}. As a basic check of the formalism, we reproduce the known leading-order result for relativistic dynamical friction~\cite{Barausse:2007ph} from a tree-level amplitude for single phonon emission.

Another important environmental effect is accretion, in which the compact object absorbs matter together with its energy and momentum, providing a distinct dissipative channel that has been studied extensively in both nonrelativistic and relativistic settings~\cite{Bondi:1944jm, Bondi:1952ni, Michel:1972oeq, Petrich:1989}. Although we ignore such short-distance effects in the present paper, we note that accretion can be incorporated into the EFT description of compact objects with localized worldline response operators~\cite{Goldberger:2005cd, Wong:2019yoc, Goldberger:2020wbx}. Although it clearly applies to one-body environmental effects, this EFT formalism is a natural framework for extending post-Minkowskian computations of binary dynamics.

The EFT framework describes systems where the wavelengths and orbital scales of interest are well separated from the short-distance length scales, as is often the case for environmental effects on a compact object. Rather than needing to specify the short-distance dynamics, an EFT identifies the relevant long-distance degrees of freedom, constructs their interactions from the symmetries of the system, and uses power counting to determine the terms required at a given accuracy. For a relativistic fluid, the low-energy degrees of freedom can be identified as sound mode fluctuations on a homogeneous background, which are referred to as phonons~\cite{Dubovsky:2005xd, Endlich:2010hf}.
This symmetry-first, field-theoretic formulation of hydrodynamics is well-studied in flat spacetime~\cite{Dubovsky:2011sj, Gripaios:2014yha}, has been extended to incorporate dissipative and stochastic channels using the Schwinger-Keldysh formalism~\cite{Liu:2018kfw} (see also Ref.~\cite{Esposito:2026diu} for a recent review), and has been covariantly coupled to gravity to study non-vacuum and cosmological spacetimes (see, e.g., Refs.~\cite{Ballesteros:2012kv, Ballesteros:2014sxa, Celoria:2017bbh, Modrekiladze:2024htc}). This class of effective theories has even been used to model inflation~\cite{Endlich:2012pz, Gruzinov:2004ty} and superfluid dark matter~\cite{Acanfora:2019con, Berezhiani:2019pzd}, among other systems.

Several applications of this field-theoretic formulation of hydrodynamics incorporate the ingredients needed for systematic calculations of environmental corrections to binary dynamics. In gravitational settings, related effective theories have been used to calculate Newtonian dynamical friction in superfluid dark matter models~\cite{Berezhiani:2019pzd, Berezhiani:2023vlo}. In quantum chromodynamics, similar theories describe energetic jets traversing a quark-gluon plasma at colliders, providing a framework for calculating observable energy and momentum transfer to a dynamical medium through non-Abelian gauge interactions~\cite{Cao:2020wlm, Kirchner:2025xdb}. Field-theoretic treatments of relativistic media have derived gravitational drag at leading order from the linear response of the medium encoded in its stress-tensor correlator~\cite{Katz:2019rgf}, and relativistic Einstein-fluid perturbation theory has been developed to study gravitational-wave generation and propagation in nonvacuum black-hole spacetimes~\cite{Cardoso:2022whc}. Beyond leading order, the nonlinear response of the medium enters, and for a fluid it is governed by the self-interactions of its phonons and their nonlinear couplings to gravity. What has been missing is a formulation in which gravitons, phonons, and the compact object appear as fields with local interaction vertices and standard propagators. The nonlinear response is then generated by Feynman diagrams, and higher-order contributions can be built and evaluated using generalized unitarity and multiloop integration methods from post-Minkowskian calculations. We construct such an EFT, derive its Feynman rules valid through four points, and verify them with generalized Ward identities. For on-shell amplitudes, the diagrams in our setup satisfy a selection rule that sets to zero any diagram that mixes an on-shell transverse-traceless (TT) graviton, in the fluid rest frame, with a phonon.

Gravitational on-shell Ward identities~\cite{Weinberg:1964ew} provide an important consistency condition for new Feynman rules. In vacuum, these identities express the decoupling of longitudinal graviton polarizations from amplitudes with physical external states. The homogeneous fluid background, however, selects a preferred rest frame and breaks boost invariance. Spatial diffeomorphisms consequently act inhomogeneously on the phonon Goldstone mode, leading to kinetic mixing of the phonon and graviton.  By expanding perturbatively in this mixing, we can use non-interacting phonon and graviton external states, allowing us to reuse flat-space hydrodynamics and vacuum-gravity Feynman rules to calculate compact binary dynamics in a medium.
In this basis, the resulting generalized Ward identity relates the divergence of an amplitude with an external graviton to the corresponding amplitude with that graviton replaced by a phonon. This is the gravitational analog of the Goldstone or St\"uckelberg relation in a spontaneously broken gauge theory~\cite{Ruegg:2003ps, Cuomo:2019siu, Kribs:2022gri}.  
Here we use these Ward identities to explicitly confirm the derived nonlinear fluid-gravity interaction vertices through four points. 

The remainder of this paper is organized as follows. In \autoref{sec:Hydrodynamics}, we review the covariant EFT of a perfect fluid, establish the relevant scale hierarchies, and derive the quadratic graviton-phonon action. In \autoref{sec:FeynmanRules}, we perturbatively expand the Lagrangian, define the physical states, and collect the propagators and interaction vertices arising from expanding the Lagrangian.  In \autoref{sec:ScatteringAmplitudes}, we collect scattering amplitudes with one- and two-phonon emissions. In \autoref{sec:GenWardId}, we write down the generalized on-shell Ward identities and verify them explicitly at three points. In \autoref{sec:DF_from_Amplitudes}, we demonstrate a simple calculation in this EFT, extracting the leading-order dynamical-friction force from the tree-level single-phonon emission amplitude. We collect our conventions, interaction terms, four-point vertices, and four-point Ward-identity checks in the appendices.  Accompanying ancillary files contain the explicit four-point interaction vertices and the Ward identity checks.  Our conclusions and outlook are given in \autoref{sec:Conclusions}.

\vskip .3 cm 
\centerline{\bf Note Added}
While this manuscript was being completed, complementary related work appeared~\cite{Modrekiladze:2026twz, Modrekiladze:2026drr}. Ref.~\cite{Modrekiladze:2026twz} constructs an in-in (Schwinger--Keldysh) worldline effective theory for a compact object interacting with a general environment through its stress-tensor correlators, including dissipative and stochastic effects. It derives the leading-order dynamical friction in dust, fluids, and other media, and analyzes the renormalization-group running and short-distance matching of the drag coefficient. Ref.~\cite{Modrekiladze:2026drr} builds on this effective theory of gravity in a relativistic fluid with the medium integrated out, including three-point fluid-gravity couplings and the medium's gravitational response.  In contrast to those papers, we keep the phonon as an explicit field coupled to gravity. Moreover, they treat dissipation and short-distance effects, which we have not attempted here and leave for the future.

\section{Curved space hydrodynamic EFT}
\label{sec:Hydrodynamics}

\subsection{Basic setup}

Over the last twenty years, there has been considerable interest in creating a field-theoretic description of hydrodynamics, following the EFT program to write down the most general possible fluid from symmetry considerations and power counting alone.  The first studies developed the leading nontrivial description in flat space, imposing the basic internal symmetries of an ideal fluid to build up an effective action for fluid fluctuations \cite{Dubovsky:2005xd, Endlich:2010hf, Dubovsky:2011sj, Gripaios:2014yha, Nicolis:2013lma}.  More recent work has extended this framework to curved space~\cite{Bhattacharya:2012zx, Aoki:2022ipw, Ballesteros:2012kv, Celoria:2017bbh}, reformulated in Schwinger-Keldysh to include channels for dissipation \cite{Crossley:2015evo, Glorioso:2017fpd, Haehl:2015uoc, Jensen:2017kzi, Liu:2018kfw}, and classified symmetry-allowed operators in the non-dissipative action at next-to-leading and higher orders in the derivative expansion
\cite{Bhattacharya:2012zx, Ballesteros:2014sxa, Aoki:2022ipw}. 

In this section, we review how to construct the effective theory for a perfect inviscid fluid in general relativity. We follow the conventions listed in Appendix \ref{app:conventions}, expanding the metric about flat space as $g_{\mu\nu} = \eta_{\mu\nu} + \kappa h_{\mu\nu}$ for $\kappa^2 = 32 \pi G$, though the EFT is still valid in strongly curved backgrounds. As is typical in the EFT of an inviscid fluid in flat spacetime of dimension $D$ \cite{Endlich:2010hf, Dubovsky:2011sj}, we describe the fluid with $D-1$ scalar comoving coordinates, one for each spatial degree of freedom:\footnote{Although our primary interest is in four-dimensional applications, higher-order calculations employing dimensional regularization are naturally formulated in $D$ dimensions.}
\begin{equation}
\phi^i = \phi^i(t,\mathbf{x}), \qquad i = 1,\dots, D-1 .
\end{equation}
The homogeneous fluid at rest is the state
\begin{equation}
\langle \phi^i(t,\mathbf{x}) \rangle = x^i \quad \forall \, t .
\end{equation}
Phonons are fluctuations about the comoving background coordinate,
\begin{equation}
\phi^i = x^i + \lambda \pi^i(t,\mathbf{x}),
\label{eq:phi-expansion}
\end{equation}
with $\lambda$ a counting parameter for the number of phonon fields.  We then expand about this homogeneous reference state with no fluid or gravity fluctuations.  

Besides spacetime symmetries, the low-energy action is invariant under internal translations, rotations, and volume-preserving diffeomorphisms,
\begin{subequations}
\label{eq:fluid_symmetries}
\begin{align}
\phi^i &\to \phi^i + a^i, \qquad a^i = \text{constant}, \\
\phi^i &\to R^i_{\ j}\phi^j, \qquad R \in \mathrm{SO}(D-1), \\
\phi^i &\to \xi^i(\phi), \qquad
\det\!\left(\frac{\partial \xi^i}{\partial \phi^j}\right)=1 .
\end{align}
\end{subequations}
In integer dimension $D$, these symmetries imply the identically conserved entropy current, $\nabla_\mu \mathcal{J}^\mu=0$, for 
\begin{equation}
\mathcal J^\mu
=\frac{1}{(D-1)!}\,\epsilon^{\mu \mu_1 \dots \mu_{D-1}}\,
\tilde\epsilon_{i_1 \dots i_{D-1}}\,
\partial_{\mu_1}\phi^{i_1}\,\dots\,\partial_{\mu_{D-1}}\phi^{i_{D-1}}
=\frac{J_\text{flat}^\mu}{\sqrt{|g|}},
\end{equation}
where 
\begin{equation}
J_\text{flat}^\mu
= \frac{1}{(D-1)!}\,\tilde\epsilon^{\mu \mu_1 \dots \mu_{D-1}}\,
\tilde\epsilon_{i_1 \dots i_{D-1}}\,
\partial_{\mu_1}\phi^{i_1}\,\dots\,\partial_{\mu_{D-1}}\phi^{i_{D-1}}
\end{equation}
is the entropy current in flat space. 
We distinguish the numerical antisymmetric symbol $\tilde\epsilon^{\mu_1 \dots \mu_D}$, which is a tensor density, from the true Levi-Civita tensor,
\begin{align}
\epsilon^{\mu_1 \dots \mu_D}
&= \frac{1}{\sqrt{|g|}}\;\tilde\epsilon^{\mu_1 \dots \mu_D}.
\end{align}
Continuing the Levi-Civita tensor to noninteger dimensions in dimensional regularization requires care, as briefly discussed in
\autoref{sec:feynman_rules}. To avoid these subtleties, here we take
$D\in\mathbb{Z}$, with the sign convention $\tilde \epsilon^{01\cdots D-1}=+1$. Using \eqn{eq:phi-expansion}, the flat-space entropy current expands as
\begin{subequations}
\begin{align}
J^0_\text{flat}
&= 1 + \lambda \partial_i \pi^i +\frac{\lambda^2}{2}\Big[(\partial_i \pi^i)^2 - \partial_i \pi^j \partial_j \pi^i \Big] + \frac{\lambda^3}{6}\Big[(\partial_i \pi^i)^3 - 3 \partial_i \pi^i \partial_j \pi^k \partial_k \pi^j + 2 \partial_i \pi^j \partial_j \pi^k \partial_k \pi^i \Big] + \cdots,\\
J^i_\text{flat}
&= - \lambda \dot\pi^i + \lambda^2\Big[\dot \pi^j \partial_j \pi^i - \dot \pi^i (\partial_j \pi^j)\Big] +\lambda^3 \Bigg[-\frac{1}{2}\left((\partial_j \pi^j)^2 - \partial_j \pi^k \partial_k \pi^j\right)\dot{\pi}^i + (\partial_j \pi^j)\dot{\pi}^k \partial_k \pi^i - \dot{\pi}^j \partial_j \pi^k \partial_k \pi^i\Bigg] + \cdots.
\end{align}
\end{subequations}
This finite expansion terminates at order $\lambda^{D-1}$.

The Lorentz invariant entering the leading hydrodynamic action is therefore
\begin{align}
X
&
= g_{\mu\nu} \mathcal J^\mu \mathcal J^\nu
= \frac{g_{\mu\nu} J_\text{flat}^\mu J_\text{flat}^\nu}{|g|} = (-1)^{D-1} \det B,
\end{align}
for $B^{ij} = g^{\mu\nu} \partial_\mu \phi^i \partial_\nu \phi^j$. The final equality is the standard Gram-determinant identity in a mostly minus metric, where the Levi-Civita tensors in the contraction $\mathcal J^\mu \mathcal J_\mu$ antisymmetrize the gradients and produce the determinant of their pairwise contractions $B^{ij}$. 
The zeroth order in the derivative expansion is a general function of this Lorentz invariant, minimally coupled to gravity,
\begin{equation}
\begin{aligned}
S \supset -\int \mathrm{d}^Dx\;\sqrt{|g|}\,w_0 f(b)
\end{aligned}.
\end{equation}
where, following standard treatments of the EFT for flat space hydrodynamics,\footnote{We note that in some of the literature (e.g. \cite{Endlich:2010hf}), the action is instead written in terms of the generic function $F(X) = -w_0 \,f(b).$} we factored out the scale dependence of this general function with
\begin{align}
\qquad b\equiv\sqrt X,
\qquad [w_0]=D,
\qquad [f]=0.
\end{align}
Corrections to the pure fluid part of the EFT begin only at second order in the derivative expansion \cite{Bhattacharya:2012zx}, which is briefly discussed in \autoref{sec:scaling}.
The freedom to rescale $w_0$ and $f$ is fixed by setting $f'(1)=1$. The energy-momentum tensor of the fluid is obtained by variation of the fluid term with respect to the metric.
\begin{equation}
\label{eq:f_derived_stress_tensor}
\begin{aligned}
T^{\mu\nu}
&= w_0\left[f(b)g^{\mu\nu}-b f'(b)(B^{-1})^{ij} \,
\partial^\mu \phi^j \partial^\nu \phi^i\right].
\end{aligned}
\end{equation}
With the mostly-minus metric signature $({+}{-}\;{\cdots}\;{-})$, the uncharged perfect-fluid stress tensor is
\begin{align}
\label{eq:perfect_fluid_stress_tensor}
T^{\mu\nu}
= (p+\varepsilon)u^\mu u^\nu - p\,g^{\mu\nu}.
\end{align}
Here $u^\mu = \mathcal J^\mu / b$ is the fluid $D$-velocity normalized with $u^\mu u_\mu = 1$, while $\varepsilon$ and $p$ are the fluid energy density and pressure.  Contracting both forms of the stress tensor \eqref{eq:f_derived_stress_tensor} and \eqref{eq:perfect_fluid_stress_tensor} with $u_\mu u_\nu$ and $g_{\mu\nu}$, then using $\mathcal J^\mu \partial_\mu \phi^i = 0$ identifies the state-dependent thermodynamic functions 
\begin{subequations}
\begin{align}
\varepsilon(b)
&=w_0 f(b),\\[6pt]
p(b)
&=w_0\bigl[b f'(b)-f(b)\bigr],\\[6pt]
w(b)&\equiv\varepsilon(b)+p(b)=w_0 b f'(b).
\end{align}
\end{subequations}
The last function is the enthalpy density, defined as the sum of the energy density and pressure. The state-dependent sound speed is given by the usual thermodynamic relation,
\begin{equation}
c_s^2(b)
\equiv\frac{\mathrm d p(b)}{\mathrm d\varepsilon(b)}
=b\frac{f''(b)}{f'(b)}.
\end{equation}
Expanding about the homogeneous fluid background $b = 1$ with $D$-velocity $u^\mu = (1,\mathbf 0)$ in the fluid rest frame, the scale $w_0 = w(1)$ is identified with the rest enthalpy density of the fluid. The dimensionless Wilson coefficients of the fluid EFT are the derivatives of $f$ for a fluid at rest, $f_n \equiv f^{(n)}(1)$, which are fixed upon choosing an equation of state. The first nonlinear coefficient measures the compression dependence of the sound speed,
\begin{equation}
w_0\left.\frac{\mathrm d c_s^2(b)}{\mathrm d\varepsilon(b)}\right|_{b=1}
=c_s^2+f_3-c_s^4.
\end{equation}
Here $c_s^2$ without an argument denotes the background sound speed, which is identified with $f''(1)$ below. Higher-order Wilson coefficients are related to higher derivatives of the speed of sound.

The most general local conservative action can be written as
\begin{equation}
\begin{aligned}
\int \!\mathrm{d}^Dx \sqrt{|g|}\, \biggl[
&\frac{2}{\kappa^2}R -w_0 f(b)
+ \frac{1}{2}(\partial_\mu \chi)^2 - Z_m(b)\frac{1}{2}m^2\chi^2
+ Z_u(b)\frac12 ({\cal J}^\mu \partial_\mu \chi)^2
\biggr] + \cdots,
\end{aligned}
\label{eq:GeneralLagrangian}
\end{equation}
where $\chi$ is a scalar field of mass $m$ representing the compact object, and we have imposed $\mathbb Z_2$ symmetry to rule out the scalar $\mathcal J^\mu \partial_\mu \chi$.  
We work in a field basis where the Einstein-Hilbert term and the scalar field kinetic term take the canonical forms, and the ellipsis denotes higher-derivative operators like $R_{\mu\nu}\mathcal{J}^\mu \mathcal{J}^\nu$.
The functions $Z_m(b)$ and $Z_u(b)$ include nonminimal coupling of the compact object to the medium, analogous to higher-order response data in phonon EFTs \cite{Acanfora:2019con, Matchev:2021fuw}. These operators describe finite-size effects and are suppressed by the point-particle expansion parameter, as in the case of tidal operators.  Universal long-range effects such as dynamical friction are isolated by taking the minimally coupled limit $Z_m (b)=1$ and $Z_u(b)=0$, and neglecting $b$-derivatives of these functions.  This choice removes direct phonon-worldline contact terms dependent on short-distance matching, while preserving the gravitationally mediated phonon emission that reproduces known superfluid and inviscid-fluid results \cite{Berezhiani:2019pzd, Berezhiani:2023vlo}.  The action used below is therefore
\begin{equation}
S = \int \!\mathrm{d}^Dx \sqrt{|g|}\, \biggl[
\frac{2}{\kappa^2}R -w_0 f(b)
+ \frac{1}{2}(\partial_\mu \chi)^2 -\frac{1}{2}m^2\chi^2
\biggr] \, . 
\label{eq:CleanedAction}
\end{equation}
For many-body dynamics, there will be a massive scalar field $\chi_i$ for each compact object. The pure gravity Einstein-Hilbert action requires gauge fixing.
We follow the standard procedure to add a gauge-fixing term to the Lagrangian, 
\begin{equation}
S_{hh}^{\rm GF} = \int \! \mathrm{d}^D x \, 
(\partial^\mu h_{\mu\nu}-\tfrac12\partial_\nu h)^2
  \,.
  \label{eq:hhGaugeFix}
\end{equation}
In the quantum theory, Faddeev-Popov ghost terms are needed, but we can ignore them in the classical theory, as we do below.

Note that here we focus on the purely conservative terms. The most general action includes genuinely dissipative channels and counterterms, and must be written in a doubled-field (Schwinger-Keldysh) formalism (see, e.g., Refs.~\cite{Crossley:2015evo, Modrekiladze:2026twz}).  

\subsection{Scales and power counting}
\label{sec:scaling}

\begin{figure}[tb]
\centering
\ManuscriptFigurePDF[scale=1]{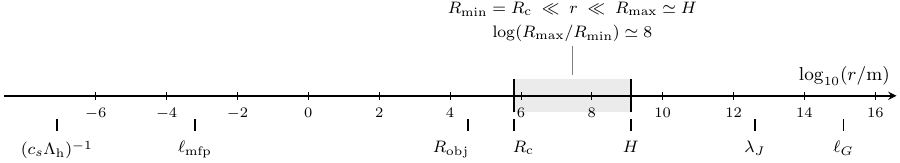}%
\caption{The scales of \autoref{tab:scales} for the thin-disk EMRI of Eq.~\eqref{eq:benchmark}. The wake at distance $r$ is resolved by fluctuations of momentum $K\sim1/r$. The EFT applies in the shaded window, given by Eq.~\eqref{eq:window}. It is bounded below by the capture radius $R_\text{c}$ of the perturber and above by the disk scale height $H$, which sets the size $R_\text{fluid}$ of the homogeneous region. The Coulomb logarithm of the dynamical-friction force is the logarithmic width of the window. The mean free path, the hydrodynamic scale, and the de~Broglie wavelength lie far below, and the Jeans length $\lambda_J=c_s\ell_G$ and the mixing scale $\sqrt{1-c_s^2}\,\ell_G\simeq\ell_G$ far above.}
\label{fig:scales}
\end{figure}

The problem involves three physical systems: the fluid, the perturber, and gravity. Each one has its own characteristic lengths, and the EFT applies in the window of distances over which all three are under perturbative control. Environmental effects can be important for EMRIs with a thin accretion disk~\cite{Barausse:2014tra}, making it a useful benchmark case. 
\autoref{tab:scales} lists the scales involved and \autoref{fig:scales} shows them for a representative thin-disk system. We work in the fluid rest frame, where the perturber has momentum $p^\mu=\gamma m(1,\mathbf v)$ with Lorentz factor $\gamma = 1/\sqrt{1-|\mathbf v|^2}$. We denote by $k^\mu=(\omega,\mathbf k)$, $K\equiv|\mathbf k|$, the momentum of a bulk fluctuation, graviton or phonon. The wake at a distance $r$ from the perturber is built from fluctuations of momentum $K\sim 1/r$, since it is sourced by the perturber's gravitational field at that distance. The window where the EFT applies can therefore be read either as bounds on $K$ or as bounds on the distance $r\sim1/K$.

\begin{table}[tb]
\centering
\renewcommand{\arraystretch}{1.5}\setlength{\tabcolsep}{6pt}
\begin{tabular}{@{}l l l r@{}}
\hline\hline
Scale & Requirement & Condition & Benchmark \\
\hline
$p^{-1}$ & classical (eikonal) perturber & $r\gg p^{-1}$ & $10^{-73}$ m \\
$(c_s\Lambda_\text{h})^{-1}$ & linear phonon dispersion & $r\gg (c_s\Lambda_\text{h})^{-1}$ & $10^{-7}$ m \\
$\ell_\text{mfp}$ & hydrodynamic description of the medium & $r\gg\ell_\text{mfp}$ & $7\times10^{-4}$ m \\
$R_\text{obj}$ & point-particle perturber & $r\gg R_\text{obj}$ & $3\times10^{4}$ m \\
$R_\text{c}$ & small deflection of fluid elements ($R_\text{c}=Gm(1+v^2)/v^2$) & $r\gg R_\text{c}$ & $6\times10^{5}$ m \\
\noalign{\hrule height 1.2pt}
$R_\text{fluid}$ & homogeneous medium (here the disk scale height $H$) & $r\ll R_\text{fluid}$ & $1\times10^{9}$ m \\
$\lambda_J=c_s\ell_G$ & Jeans stability, linear dispersion & $r\ll\lambda_J$ & $4\times10^{12}$ m \\
$\sqrt{1-c_s^2}\,\ell_G$ & perturbative mixing ($\ell_G=(4\pi Gw_0)^{-1/2}$) & $r\ll\sqrt{1-c_s^2}\,\ell_G$ & $1\times10^{15}$ m \\
\hline\hline
\end{tabular}
\caption{The length scales of the theory, the requirement each encodes, and the condition it imposes on the distance $r\sim1/K$ at which the fluid wake is resolved. The scales above the thick line bound the window Eq.~\eqref{eq:window} from below, $R_\text{min}=\max(\cdots)$, while those below it bound it from above, $R_\text{max}=\min(\cdots)$. The Benchmark column evaluates them for the thin-disk extreme-mass-ratio inspiral of Eq.~\eqref{eq:benchmark}. The window then spans $R_\text{min}=R_\text{c}$ to $R_\text{max}\simeq H$, with $\log(R_\text{max}/R_\text{min})\simeq8$.}
\label{tab:scales}
\end{table}

\begingroup
\renewcommand{\addcontentsline}[3]{}
\subsubsection*{The fluid medium}
The mean free path $\ell_\text{mfp}$ of the constituents of the medium must be shorter than any distance resolved, $r\gg\ell_\text{mfp}$. This is the condition for a hydrodynamic description, and, since viscous corrections are suppressed by $\ell_\text{mfp}/r$, it is also what justifies treating the fluid as inviscid.
Beyond this, the hydrodynamic EFT is a derivative expansion: operators with more derivatives are suppressed by a hydrodynamic scale $\Lambda_\text{h}\sim w_0^{1/D}$, and because the entropy current is identically conserved the first bulk corrections to the action \eqref{eq:CleanedAction} arise at second order, with the structure $w_0\big[c_2\,\nabla_\mu b\nabla^\mu b + c_3\,\nabla_\mu\mathcal J_\nu\nabla^\mu\mathcal J^\nu + c_4\, R+\cdots\big]/\Lambda_\text{h}^2$, where the $c_i$ are functions of $b$ (see Ref.~\cite{Bhattacharya:2012zx} for the complete list). These operators correct the phonon dispersion relation to $\omega^2=c_s^2K^2+\mathcal O(K^4/\Lambda_\text{h}^2)$. Throughout, we assume linear dispersion, which for $0<c_s<1$ requires $K\ll c_s\Lambda_\text{h}$. The fluid is therefore under control for $r\gg\max\big(\ell_\text{mfp},\,(c_s\Lambda_\text{h})^{-1}\big)$.

\subsubsection*{The perturber as a point mass in weak gravity}
We expand the metric about flat space, $g_{\mu\nu}=\eta_{\mu\nu}+\kappa h_{\mu\nu}$ with $\kappa^2=32\pi G$, and the fluid about its homogeneous state, $\phi^i=x^i+\lambda\pi^i$ with $\lambda=1/\sqrt{w_0}$ (\autoref{sec:quadratic_terms}). After canonical normalization, the two counting parameters have mass dimension
\begin{equation}
\label{eq:counting_param_dims}
[\kappa] = (2-D)/2, \hskip 1cm [\lambda] = -D/2\,,
\end{equation}
and for loop-level applications in $D=4-2\epsilon$, we identify $[\kappa]=-1$, $[\lambda]=-2$, introducing a renormalization scale $\mu$ as usual. Three conditions make the perturber a classical point mass in weak gravity. First, it is classical when $r$ exceeds its de~Broglie wavelength, $r\gg p^{-1}$ with $p=\gamma m|\mathbf v|$, which also places every exchange in the eikonal regime $\omega,K\ll m$ familiar from the extraction of classical post-Minkowskian physics from amplitudes~\cite{Cheung:2020gyp, Bern:2019nnu, Kosower:2018adc}. Second, it is point-like when the fluctuations do not resolve the object's size, $r\gg R_\text{obj}$.\footnote{Here $R_\text{obj}$ is the length scale of the perturber in its own rest frame. No Lorentz factor enters the bound. In contrast, a static inhomogeneity of the medium would be Lorentz contracted in the perturber frame, as in Ref.~\cite{Modrekiladze:2026twz}.} Finite-size effects are suppressed by powers of $K R_\text{obj} \sim R_\text{obj}/r$.
Third, gravity is weak. The metric perturbation at distance $r$ is of order $Gm/r$, but the expansion of the wake, and hence medium-induced observables, in powers of $G$ is controlled by a larger quantity, the deflection $\theta$ of a fluid element passing the perturber at impact parameter $b$,
\begin{equation}
\label{eq:capture-radius}
    \theta\simeq\frac{2R_\text{c}}{b}\,,
    \qquad
    R_\text{c}\equiv\frac{Gm\,(1+v^2)}{v^2}\, .
\end{equation}
Here $\theta$ is, up to the relativistic factor $(1+v^2)$, the ratio of the potential energy of the fluid element to its kinetic energy, exactly as in the post-Minkowskian expansion of the vacuum scattering angle. Small deflection requires $r\gg R_\text{c}$, where $R_\text{c}$ is the relativistic generalization of the Newtonian capture radius $Gm/v^2$; the impact parameter is transverse to the motion and hence frame independent, and the full velocity dependence of the deflection is contained in $(1+v^2)/v^2$. For $v<1$ this is larger than the Schwarzschild radius $R_\text{s}=2Gm$, so for a compact perturber it is the more restrictive condition; the two coincide only as $v\to1$.  In the pressureless limit, where the one-body drag force is well understood~\cite{Vicente:2022ivh, Traykova:2023qyv},  $R_\text{c}$ is precisely the scale at which the Coulomb logarithm of the long-range dynamical friction force is cut off, and Ref.~\cite{Modrekiladze:2026twz} identifies it as the scale at which the drag coefficient is matched to the short-distance geodesic motion of the medium. For a fluid with pressure, the corresponding scale is the accretion radius $\sim Gm/(v^2+c_s^2)$, which coincides with $R_\text{c}$ up to an $\mathcal O(1)$ factor in the supersonic regime $v>c_s$ and is smaller than $R_\text{c}$ for subsonic motion. Since $R_\text{c}$ is the more restrictive of the two in either regime, we use it in the window of scales below. Altogether, the perturber is under control for $r\gg\max\big(p^{-1},\,R_\text{obj},\,R_\text{c}\big)$.

\subsubsection*{Gravity as a weak perturbation of the fluid}
The self-gravity of the medium introduces one more length,
\begin{equation}
    \ell_G\equiv\frac{1}{\sqrt{4\pi G w_0}}=\frac{2\sqrt2\,\lambda}{\kappa}\,,
\end{equation}
the curvature radius of the medium. A homogeneous region of size $R$ has compactness $GM_\text{fluid}/R\sim (R/\ell_G)^2$ for $M_\text{fluid}\sim w_0R^3$, so $\ell_G$ is the size at which its self-gravity becomes strong. Equivalently, $\ell_G$ is the distance light travels in the gravitational free-fall time $(4\pi Gw_0)^{-1/2}$ of the medium. The Jeans length $\lambda_J=c_s\ell_G$ introduced below is the shorter distance that sound travels in the same time. 

The length $\ell_G$ controls the graviton-phonon mixing of the quadratic action \eqref{eq:KineticGravitonPhonon}. Relative to the kinetic terms, which scale as $K^2$, the $h\pi$ mixing scales as $(\kappa/\lambda)K$, so each mixing insertion costs a factor $(\kappa/\lambda)/K\sim r/\ell_G$. This is the sense in which we ``expand perturbatively in $\kappa/\lambda$'' throughout the paper, and it is a good expansion only for $r\ll\ell_G$, i.e.\ when the fluid within a distance $r$ of the perturber is weakly self-gravitating, and the flat background is a good approximation. Two effects make this condition depend on the sound speed. First, integrating out the graviton from the quadratic action shifts the phonon pole, in $D=4$, to
\begin{equation}
\label{eq:jeans-shift}
    \omega^2=c_s^2K^2-\frac{1}{\ell_G^2}\,,
\end{equation}
the familiar Jeans term, with $\ell_G^{-2}=4\pi Gw_0$ reducing to the textbook $4\pi G\rho$ for nonrelativistic matter, where $w_0=\varepsilon_0+p_0\simeq\rho$ (see, e.g., Ref.~\cite{Berezhiani:2019pzd}).\footnote{The relativistic correction to \eqref{eq:jeans-shift} is of relative order $c_s^2$, the same order as the background-curvature terms neglected in expanding about flat space, and is dropped. Resumming the same mixing insertions on a static graviton line instead moves the pole of the Newtonian potential to $k_J=(c_s\ell_G)^{-1}$, the linearized Poisson-Jeans system (see, e.g., Ref.~\cite{Modrekiladze:2026drr}).} Modes with $K<k_J\equiv1/(c_s\ell_G)$ are unstable, so a homogeneous background exists and phonons propagate with linear dispersion only on scales much below the Jeans length: $r\ll\lambda_J=c_s\ell_G$. Second, the mixing vertex converts a phonon into a graviton of the same four-momentum. For an on-shell phonon, $\omega=c_sK$, this graviton is off shell, $k^2=-(1-c_s^2)K^2$, and its propagator is $1/[(1-c_s^2)K^2]$ rather than $1/K^2$, so a mixing insertion actually costs $r/(\sqrt{1-c_s^2}\,\ell_G)$ rather than $r/\ell_G$. Gravity is therefore a weak perturbation of the fluid for
\begin{equation}
    \label{eq:weak_mixing_condition}
    r\;\ll\;\min\!\left(c_s,\;\sqrt{1-c_s^2}\right)\ell_G\,,
    \qquad\text{equivalently}\qquad
    \frac{\kappa}{\lambda}\;\ll\;\min\!\left(c_s,\;\sqrt{1-c_s^2}\right)K\,.
\end{equation}
In this regime, pure gravitons and pure phonons remain good asymptotic states, the multiplet propagator of \autoref{sec:quadratic_terms} can be expanded in mixing insertions, and the graviton tadpole, the $\mathcal O(\kappa^2w_0h^2)$ background terms discussed there, and medium-induced corrections to the pure-gravity vertices can all be dropped. In computing observables from unitarity cuts in \autoref{sec:DF_from_Amplitudes}, the first non-vanishing graphs at leading order in $\kappa/\lambda$ typically carry a mixing insertion on each side of the cut, since phonons couple to the perturber only through gravity.

\subsubsection*{The window of scales}
Collecting the three sets of conditions, and denoting by $R_\text{fluid}$ the size of the region over which the medium can be treated as homogeneous and at rest (necessarily $R_\text{fluid}\lesssim\lambda_J$), the EFT and its perturbative expansions apply for
\begin{equation}
\label{eq:window}
\begin{aligned}
    &R_\text{min}\;\ll\; r\;\ll\; R_\text{max}\,,\\[2pt]
    &R_\text{min}=\max\!\Big(p^{-1}, \; R_\text{obj},\;R_\text{c},\;\ell_\text{mfp},\;(c_s\Lambda_\text{h})^{-1}\Big),\\[2pt]
    &R_\text{max}=\min\!\Big(R_\text{fluid},\; \lambda_J,\;\sqrt{1-c_s^2}\,\ell_G\Big).
\end{aligned}
\end{equation}
The corresponding momenta $K_\text{max}=1/R_\text{min}$ and $K_\text{min}=1/R_\text{max}$ are the scales that cut off the Coulomb logarithm of the dynamical-friction force in \autoref{sec:DF_from_Amplitudes}. The infrared end is set by the size of the medium, its Jeans length, or its self-gravity scale, and the ultraviolet end by whichever of the perturber's size, its capture radius, or the fluid's microphysics is resolved first.

For the purpose of illustrating a region of validity for our EFT setup in a concrete example, we choose a geometrically thin Shakura-Sunyaev accretion disk model~\cite {Shakura:1972te}, following the discussion in Ref.~\cite{Barausse:2014tra}. For a black hole moving at nonrelativistic speed through a gas,\footnote{The gas itself is an excellent ideal fluid on the scales of the wake, with Reynolds number $r/\ell_\text{mfp}\gtrsim10^{9}$.} the standard choice in the astrophysical literature is $R_\text{min}=R_\text{c}$ and $R_\text{max}=R_\text{fluid}$. \autoref{tab:scales} and \autoref{fig:scales} make this concrete for a thin-disk extreme-mass-ratio inspiral, the environment in which dynamical friction is expected to be most relevant~\cite{Barausse:2014tra}. The benchmark is a $m=10\, M_\odot$ black hole on a circular orbit of radius $r_\text{orbit}=40\, GM/c^2$ around a $M=10^6\, M_\odot$ black hole, so that $v=v_K=(GM/c^2r_\text{orbit})^{1/2}=0.16$, in a geometrically thin, Shakura-Sunyaev accretion disk with Eddington ratio $f_\text{Edd}=\dot M/\dot M_\text{Edd}=0.1$ and viscosity parameter $\alpha=0.1$.\footnote{The viscosity parameter $\alpha$ of the disk model parametrizes turbulent transport by eddies of size $\sim H$, whose effect on the wake at $r\ll H$ is a stochastic perturbation of relative size $\sim\alpha^{1/2}(r/H)^{1/3}$, at most $\sim0.3$ near $r\sim H$: an $\mathcal O(1)$ uncertainty in the effective infrared cutoff that does not affect the hierarchy of scales and enters the drag force only through the Coulomb logarithm.} Following Ref.~\cite{Barausse:2014tra}, we take the velocity relative to the gas to be $v_K$, appropriate for an orbit inclined at $\iota=60^\circ$ to the disk. Such an orbit crosses the disk twice per period and is inside it only $1.5\%$ of the time, so the force of \autoref{sec:DF_from_Amplitudes} is the instantaneous force during a crossing.\footnote{Repeated interactions with the disk typically changes the inclination and eccentricity over many orbits, as discussed in Refs. \cite{Spieksma:2025wex,Zeng:2026ydj}. An orbit embedded in the disk with small eccentricity $e$ moves relative to the gas at $v\simeq e\,v_K$ at all times; see Ref.~\cite{Duque:2024mfw}.}

For this benchmark, the disk is radiation-pressure dominated, and the standard one-zone $\alpha$-disk solution~\cite{Shakura:1972te} with gas and radiation pressure gives, at the orbit,
\begin{equation}
\label{eq:benchmark}
\rho\simeq7\times10^{-8}\,\text{g\,cm}^{-3},\qquad
H\simeq0.8\,\frac{GM}{c^2}=1.2\times10^{9}\,\text{m},\qquad
c_s\simeq3\times10^{-3},
\end{equation}
a Mach number $v/c_s\simeq50$. Radiation is trapped only on scales above $\sim4\times10^{7}$ m; below that, the wake sees the sound speed of the gas alone, $c_s\simeq4\times10^{-4}$ (Mach number $\simeq400$). Either way, the motion is highly supersonic, and the leading-order force of \autoref{sec:DF_from_Amplitudes} does not depend on $c_s$ beyond that. The remaining entries of \autoref{tab:scales} follow from these: $\ell_\text{mfp}$ is the Coulomb mean free path of the ionized gas at the midplane temperature $T\sim7\times10^5$~K, $\Lambda_\text{h}\sim w_0^{1/4}$, and $\ell_G$ and $\lambda_J$ follow from $\rho$.

The window spans some three decades, from the capture radius of the perturber to the scale height of the disk,\footnote{The scale height is the infrared cutoff because the homogeneous, three-dimensional wake described here exists only for $r\lesssim H$. On larger in-plane scales the perturbation becomes two-dimensional, and the differential rotation and radial structure of the disk enter, physics that the astrophysical literature treats as disk torques or migration~\cite{Barausse:2014tra,  HegadeKR:2025dur, Duque:2025yfm,  Dyson:2026ddd, Dittmann:2026rtj} and that would require expanding the fluid EFT about a rotating, inhomogeneous background.} and every other scale in Eq.~\eqref{eq:window} lies many orders of magnitude outside it. The window closes in both limits of the sound speed. As $c_s\to0$ the ultraviolet bound $(c_s\Lambda_\text{h})^{-1}$ rises while the infrared bound $c_s\ell_G$ falls. At $c_s=0$ the leading action has no propagating sound mode, and a higher-gradient term must be promoted before formulas with an external on-shell phonon can be continued to that endpoint. As $c_s\to1$ the phonon and graviton poles become degenerate, $\sqrt{1-c_s^2}\,\ell_G\to0$, and the full quadratic kernel must be diagonalized; for the steady-state dynamical friction of \autoref{sec:DF_from_Amplitudes}, which requires supersonic motion $v>c_s$, this limit is never reached.
\endgroup

\subsection{Quadratic terms in action}
\label{sec:quadratic_terms}

After simplifying the quadratic terms in the above action \eqref{eq:CleanedAction} together with the gauge fixing \eqref{eq:hhGaugeFix}, the phonon-graviton kinetic terms in the action can be written as
\begin{align}
S_{\rm kin} = \int \!\mathrm{d}^Dx  \biggl[ -\frac{1}{2}  h_{\mu\nu}  \tilde P^{\mu\nu,\rho\sigma} \Box h_{\rho\sigma}
     + \frac{1}{2} \pi^i\left(-\partial_0^2 \delta_{ij}+ c_s^2 \partial_i \partial_j \right)\pi^j
    + \frac{\kappa}{2 \lambda}\,   \pi^i\left(
      \partial_i h_{00} -2\partial_0 h_{0i}
      + c_s^2\partial_i h_{jj} \right)
\biggr]\,,
\label{eq:KineticGravitonPhonon}
\end{align}
where we used integration by parts and dropped corrections to the kinetic terms, as explained in \autoref{sec:scaling}. Flat space is not an exact solution in the presence of the background fluid: expanding Eq.~\eqref{eq:CleanedAction} also produces a graviton tadpole $\propto\kappa\,T_0^{\mu\nu}h_{\mu\nu}$ and mass-like terms of order $\kappa^2w_0h^2$. The tadpole signals that the homogeneous medium sources a gravitational field of its own. As is standard in treatments of self-gravitating media, we assume that on the scales of interest this background field is balanced by physics outside the EFT and we expand about flat space. The self-gravity of the fluid within a distance $r$ of the perturber then enters only through the mass-like terms, which are suppressed by $r^2/\ell_G^2$ relative to the kinetic terms. Consistent with the weak-mixing condition Eq.~\eqref{eq:weak_mixing_condition}, we drop them throughout.

With the de Donder gauge-fixing term Eq.~\eqref{eq:hhGaugeFix}, the graviton kinetic tensor is,
\begin{equation}
  \tilde P_{\mu\nu,\rho\sigma}
  = \frac{1}{2} \left(
    \eta_{\mu\rho}\eta_{\nu\sigma}
    +\eta_{\mu\sigma}\eta_{\nu\rho}
    -\eta_{\mu\nu}\eta_{\rho\sigma}
  \right).
  \label{eq:KineticGravitonProjector}
\end{equation}
In four dimensions, $\tilde P$ is its own inverse on symmetric tensors.  However, for loop-level applications with dimensional regularization, the inverse is  
\begin{equation}
  P_{\mu\nu,\rho\sigma}
  = \frac{1}{2} \left(
    \eta_{\mu\rho}\eta_{\nu\sigma}
    +\eta_{\mu\sigma}\eta_{\nu\rho}
    - \frac{2}{D-2}\eta_{\mu\nu}\eta_{\rho\sigma}
  \right).
  \label{eq:FeynmanGravitonProjector}
\end{equation}
where $D=4-2 \epsilon$.

The quadratic phonon action fixes
\begin{align}
    c_s^2\equiv c_s^2(1)=f''(1),
\end{align}
so phonons have a linear dispersion relation, giving the on-shell condition $(k^0)^2 = c_s^2 \mathbf{k}^2$ for a phonon with momentum $k^\mu = (k^0, \mathbf k)$.  The canonical normalization of the quadratic phonon action relates the phonon expansion parameter to the rest enthalpy density,
\begin{equation}
w_0
= \frac{1}{\lambda^2}.
\end{equation}
Thus, by canonical normalization, the mass dimensions of the field counting parameters in integer dimension $D$ are in \eqn{eq:counting_param_dims}, with the usual introduction of a renormalization scale when using dimensional regularization for loop-level applications.

In addition, there is a matter kinetic term for the compact object,
\begin{equation}
S_{\rm kin}^{\rm matter} =\int \!\mathrm{d}^D x \, \biggl[ \frac{1}{2} \partial_\mu\chi\partial^\mu\chi -\frac{1}{2}  m^2\chi^2 \biggr]\,.
\end{equation}
where $\chi$ is a point-particle scalar perturber of mass $m$.

Because the graviton and phonons mix, it can be useful to consider them as a single  graviton-phonon multiplet, 
\begin{equation}
S_{\rm kin} =
\frac12\int {\mathrm d}^D x\, \Phi^{\dagger I}  {\mathcal K}_{IJ} \, \Phi^J ,
\label{eq:SingleMultiplet}
\end{equation}
where 
\begin{equation}
\Phi^I
\equiv
\begin{pmatrix}
h_{\mu\nu}\\[1mm]
\pi^i
\end{pmatrix},  
\hskip 2 cm 
{\mathcal K} = 
\begin{pmatrix}
-\tilde P^{\mu\nu,\rho\sigma}\Box
&
\displaystyle
\frac{\kappa}{2\lambda}
\left({\mathcal B}^\dagger\right)^{\mu\nu}{}_{j}
\\[4mm]
\displaystyle
\frac{\kappa}{2\lambda}
{\mathcal B}_i{}^{\rho\sigma}
&
\displaystyle
-\partial_0^2\delta_{ij}+c_s^2\partial_i\partial_j
\end{pmatrix}.
\label{eq:GravitonPhononMultiplet}
\end{equation}
Here $I = \{\mu\nu, i\}$, so the multiplet $\Phi$ can be thought of as a 13-component field in four dimensions, ten for the graviton and three for the phonon. 
The off-diagonal terms in $\mathcal K$ are
\begin{equation}
{\mathcal B}_i{}^{\mu\nu} =
\left(\eta^{\mu0}\eta^{\nu0} + c_s^2\sum_{a=1}^{D-1}\eta^{\mu a}\eta^{\nu a} \right) \partial_i + 2 \sum_{b=1}^{D-1}\eta^{0(\mu} \eta^{\nu)b} \delta_{bi} \,\partial_0 \,.
\label{eq:OffDiagonalMatrix}
\end{equation}
These off-diagonal terms in the action complicate the definition of physical states and the resulting Feynman rules, as we explain below.

At linear order, the diffeomorphism acts as
\begin{equation}
\delta h_{\mu\nu}
=-\partial_\mu\xi_\nu-\partial_\nu\xi_\mu,
\qquad
\delta\pi^a=-\frac{\kappa}{\lambda}\xi^a.
\label{eq:multiplet-gauge-transformation-position}
\end{equation}
In field-space notation, this corresponds to a pure-gauge direction of
\begin{equation}
R^I{}_{\rho}(p)\xi^\rho
=
\begin{pmatrix}
-(\partial_\mu\xi_\nu+ \partial_\nu\xi_\mu)
\\[2.5mm]
-\dfrac{\kappa}{\lambda}\xi^a
\end{pmatrix}.
\label{eq:FieldSpaceGaugeGenerator}
\end{equation}

Other gauge-fixing choices are possible, each with its own advantages and disadvantages.  A common choice is the unitary/comoving gauge $\phi^i = x^i$, in which the phonon degrees of freedom are eaten by the spacetime metric. For a perfect fluid, the normal two TT modes of the graviton remain massless, and the graviton also acquires two non-propagating vector modes (vorticities from transverse phonons) and one massless scalar mode that propagates subluminally (longitudinal phonon) \cite{Lin:2015cqa, Aoki:2022ipw}.  This is analogous to the unitary gauge in the Higgs mechanism, in which the unphysical Goldstone modes are eaten by the gauge fields and result in massive gauge bosons with longitudinal polarizations \cite{Higgs:1964pj}. The tensor, vector, and scalar modes in unitary gauge are useful at tree-level for physical intuition, but Feynman rules in unitary gauge are often prohibitively cumbersome for renormalization and are rarely used in loop calculations \cite{Irges:2017ztc}.  One could also choose an $R_\xi$-like gauge fixing as in the Standard Model~\cite{Peskin:1995ev}, which explicitly cancels the off-diagonal terms in the action. However, the Feynman rules will be more complicated than those presented below.

\section{Lagrangian and Feynman rules}
\label{sec:FeynmanRules}
In this section, we convert the interaction Lagrangian into a set of Feynman rules needed for computing scattering amplitudes and other physical observables.
We take a perturbative approach here, treating the $h\pi$ kinetic mixing term in \eqn{eq:KineticGravitonPhonon} as a perturbative correction, suppressed by a factor of $\kappa/\lambda$.
This implies that the physical states are the usual ones, ignoring the mixing.

Upon expanding the metric about flat space, $g_{\mu\nu} = \eta_{\mu\nu} + \kappa h_{\mu\nu}$, the three-field terms of the action \eqref{eq:CleanedAction} are
\begin{align}
\label{eq:three-point-action}
    S_{}^{(3)}
    =\int \mathrm{d}^Dx\left(
    \mathcal L_{\pi^3}+\mathcal L_{h\pi^2}+\mathcal L_{h^2\pi} +\mathcal L_{h^3} + \mathcal L_{h\chi^2}
    \right).
\end{align}
A few representative three-point interaction terms in $D=4$ include 
\begingroup
\setlength{\jot}{1pt}
\begin{subequations}
\begin{align}
\mathcal L_{h^2\pi}={}&\frac{\kappa^2}{8\lambda}\biggl\{
[\partial\pi]\Bigl(
h_{00}^2-2c_s^2h_{00}h_{ii}
-(c_s^2+f_3)h_{ii}h_{jj}
+4c_s^2h_{0i}h_{0i}-2c_s^2h_{ij}h_{ij}\Bigr)
-4(h_{00}-c_s^2h_{jj})h_{0i}\dot\pi^i
\biggr\},
\\[.7 em]
\mathcal L_{h\chi^2}={}&\frac{\kappa}{2}\biggl\{\left(\frac{h}{2}\eta^{\mu\nu}-h^{\mu\nu}\right)\partial_\mu\chi\partial_\nu\chi-\frac{h}{2}m^2\chi^2\biggr\},
\end{align}
\end{subequations}
\endgroup
where $[\partial \pi] = \partial_i \pi^i$ and $h = \eta^{\mu\nu} h_{\mu\nu}$. We collect a complete list of the three- and four-point interaction terms used to derive the Feynman vertices in \app{app:interaction_terms}.

While we do not do so here, it is worth noting that it is also possible to treat the mixing exactly by treating the graviton and phonon in terms of a single multiplet \eqref{eq:GravitonPhononMultiplet} with a matrix propagator, whose four components and expansion in diagonal propagators and mixing insertions are shown in Fig.~\ref{fig:multiplet-exact-two-point-green-function}. This perspective will be useful in \autoref{sec:GenWardId} for identifying generalized on-shell Ward identities in the non-interacting basis at leading order in weak mixing.

\subsection{Physical state conditions}
\label{sec:physical_state_conditions}

Independent of how the gravity and fluid sectors are treated, we take the massive perturber to be a scalar satisfying the on-shell condition,
\begin{equation}
p^2 = m^2.
\end{equation}
We consider only the leading eikonal limit for gravity and fluid field excitations, as explained in \autoref{sec:scaling}.

Our primary way to define the physical states is by treating the off-diagonal mixing term in the Lagrangian \eqref{eq:KineticGravitonPhonon} as a two-point interaction, suppressed by the coupling $\kappa/\lambda$.  In this way, the unperturbed physical states are those of a free theory of noninteracting gravitons and phonons, so the states satisfy the usual physical state conditions.  For a graviton, we have the usual transverse-traceless (TT) conditions on the graviton polarization tensor, 
\begin{equation}
q^2 = 0,  \qquad   q^\mu \varepsilon_{\mu \nu} = 0\,, 
\qquad
n^\mu \varepsilon_{\mu \nu} = 0\,,
\qquad  \varepsilon^\mu {}_{\!\mu} = 0\,, 
\label{eq:TT-conditions}
\end{equation}
where $n_\mu$ is a null reference momentum. The residual gauge freedom in de Donder gauge allows us to choose a different null reference momentum $n_i$ for each external graviton leg $h_{\mu_i\nu_i}(q_i)$.  One convenient choice is
\begin{align}
\label{eq:fix-residual-gauge-freedom}
n_i^\mu = 2(u\cdot q_i)u^\mu - q_i^\mu
\,
\implies
\,
u^\mu \varepsilon_{\mu\nu, (i)} &= \varepsilon_{0\nu,(i)} =0,
\end{align}
where $u^\mu = (1,\mathbf{0})$ is the $D$-velocity of the homogeneous fluid background.  Henceforth, this physical polarization is referred to as $\varepsilon_{\mu\nu,(i)}^{\mathrm{TT}}$.  The longitudinal phonons travel at the speed of sound, $c_s$, through the fluid,
\begin{equation}
 \omega^2 - c_s^2 \mathbf k^2 = 0  \,,  \qquad  e_\parallel^a = \hat k^a,
\end{equation}
while the transverse ones carry zero energy,
\begin{equation}\
\omega = 0\,,  \qquad  e_\perp^a \hat k^a = 0\,,
\end{equation}
where $\hat k \equiv \mathbf k /|\mathbf k|$. See Appendix \ref{app:conventions} for further discussion about longitudinal and transverse phonon modes.  
In the rest of the paper, we will refer to these physical states of the decoupled gravity-fluid system as the \textit{non-interacting basis}.

One can also treat the full kinetic term \eqn{eq:SingleMultiplet} exactly to define the physical states, where the field-space indices $I, J$ of the graviton-phonon multiplet in \eqn{eq:GravitonPhononMultiplet} run over the original fields, $I = \{\mu\nu, i\}$.  Because of the mixing, the polarizations are also treated as a multiplet,
\begin{equation}
\widetilde u^I(n)= \begin{pmatrix}
\widetilde\varepsilon_{\mu\nu}(n)\\[2mm] 
\widetilde e^a(n)
\end{pmatrix},
\end{equation}
where the gauge invariance acts on the polarizations 
\begin{equation}
\begin{pmatrix}
\widetilde\varepsilon_{\mu\nu}(n)\\[2mm] 
\widetilde e^a(n)
\end{pmatrix} 
\longrightarrow
\begin{pmatrix}
\widetilde\varepsilon_{\mu\nu}(n)\\[2mm] 
\widetilde e^a(n)
\end{pmatrix} 
+
\begin{pmatrix}
-(\partial_\mu\xi_\nu+ \partial_\nu\xi_\mu)
\\[2.5mm]
-\dfrac{\kappa}{\lambda}\xi^a
\end{pmatrix},
\label{eq:polarizationGauge}
\end{equation}
following the gauge symmetry Eq.~\eqref{eq:FieldSpaceGaugeGenerator}.
The polarizations defined in this equation do not satisfy the separate
diagonal on-shell conditions for an isolated graviton or an isolated phonon.
Instead, they satisfy the full non-diagonal condition
Eq.~\eqref{eq:full-onshell-condition}, so the graviton and phonon slots are tied
together already at the level of the external wavefunction. Here, the states are taken to satisfy on-shell conditions dictated by the full matrix equation
\begin{equation}
 \mathcal K_{IJ}(p_i,-p_i)\widetilde u_i^J=0\,.
\label{eq:full-onshell-condition}
\end{equation}
In the rest of the paper, we will refer to these physical states of the fully interacting gravity-fluid system as the \textit{graviton-phonon multiplet basis}.

In the weak-mixing limit $\kappa / \lambda \ll q$, we can more simply express the full propagator matrix $\mathcal G=i\mathcal K^{-1}$ and the physical states $\widetilde u^I$ of the graviton-phonon multiplet as a perturbative expansion in the non-interacting basis with kinetic mixing insertions. Allowing the non-diagonal mixing between the graviton and phonon in \eqn{eq:GravitonPhononMultiplet} to be treated as a perturbation, the graviton-phonon multiplet's quadratic operator is nearly block diagonal:
\begin{align}
\mathcal K  = \mathcal K_0 - \ii \mathcal M, 
 \hskip 1 cm  {\mathcal K_0} \equiv
\begin{pmatrix}
   \mathcal K_{hh}  &   0   \\[4mm]
         0  & \mathcal K_{\pi\pi}
\end{pmatrix},   \hskip 1 cm 
{\mathcal M} \equiv \ii
    \begin{pmatrix}
    0
    &
    \mathcal K_{h \pi}
    \\[4mm]
    \mathcal K_{\pi h}
    &
    0
    \end{pmatrix}
    =
    \frac{\ii\kappa}{2\lambda}
    \begin{pmatrix}
    0
    &
    \displaystyle
    \left({\mathcal B}^\dagger\right)^{\mu\nu}{}_{j}
    \\[4mm]
    \displaystyle
    {\mathcal B}_i{}^{\rho\sigma}
    &
    0
    \end{pmatrix}
    .
\end{align}
where  $\mathcal K_{h h}$, $\mathcal K_{\pi\pi}$, and  
${\mathcal B}_i{}^{\rho\sigma}$ 
are defined in Eqs.~\eqref{eq:GravitonPhononMultiplet} and \eqref{eq:OffDiagonalMatrix}.

In the absence of mixing, the two-point Green's function for decoupled vacuum gravitons and flat-space phonons is
\begin{align}
\mathcal{G}_0 = \ii\mathcal K_0^{-1}.
\end{align}
The multiplet's two-point Green's function is instead
\begin{align}
\label{eq:full-multiplet-def}
 \mathcal G = \ii\mathcal K^{-1} = \ii \Big( \big[1 - \ii \mathcal M \mathcal K_0^{-1}\big]\mathcal K_0\Big)^{-1}
= \ii\mathcal K_0^{-1}\big[1 - \mathcal M (\ii\mathcal K_0^{-1})\big]^{-1} = \mathcal{G}_0\big[1 - \mathcal M \mathcal{G}_0\big]^{-1}.
\end{align}
Perturbatively expanding in the off-diagonal mixing terms that are of order $\kappa / \lambda$ gives the multiplet propagator matrix in terms of the non-interacting propagators with weak mixing.
\begin{equation}
\begin{aligned}
\label{eq:multiplet_prop_expansion}
    \mathcal G &= \mathcal{G}_0 + \mathcal{G}_0 \mathcal M \mathcal{G}_0 + \mathcal{G}_0 \mathcal M \mathcal{G}_0 \mathcal M \mathcal{G}_0 + \cdots\\
    &= \mathcal{G}_0 + \mathcal{G}_0 \mathcal M \mathcal{G}_0 + \mathcal O\left(\left(\kappa/\lambda\right)^2\right).
\end{aligned}
\end{equation}
Some representative matrix elements in this perturbative expansion are shown diagrammatically in \autoref{fig:multiplet-exact-two-point-green-function}.

The form of Eq.~\eqref{eq:multiplet_prop_expansion} also makes clear the perturbative expansion of the physical external states $\widetilde u^I$ satisfying the full non-diagonal on-shell condition Eq.~\eqref{eq:full-onshell-condition}.  For any null reference momentum $n_\mu$, the pure external states in a non-interacting theory, 
\begin{equation}
u_h^I(n)= \begin{pmatrix}
\varepsilon_{\mu\nu}(n)\\[2mm] 
0
\end{pmatrix},
\hskip 1 cm
u_\pi^I(n)= \begin{pmatrix}
0\\[2mm] 
e^a(n)
\end{pmatrix},
\end{equation}
satisfy the block-diagonal on-shell conditions,
\begin{align}
\label{eq:nonint_onshell_conditions}
    (\mathcal K_0 )_{IJ} u_{h}^J = 0,
    \hskip 1 cm   
    (\mathcal K_0 )_{IJ} u_{\pi}^J = 0.
\end{align}
Expanding the physical mixed external wavefunction in terms of these decoupled states,
\begin{align}
    \widetilde u_h^I(n) = u_h^I(n) + \delta u_h^I(n),
    \hskip 1 cm
    \widetilde u_\pi^I(n) = u_\pi^I(n) + \delta u_\pi^I(n),
\end{align}
the full on-shell condition for $\varphi \in \{h, \pi\}$ is given by
\begin{align}
    0 &= \mathcal K_{IJ} \widetilde u_{\varphi}^J = (\mathcal K_0 - \ii \mathcal M)_{IJ}(u_\varphi + \delta u_\varphi)^J = -\ii\mathcal M_{IJ}u_\varphi^J + \mathcal K_{IJ}\delta u_\varphi^J.
\end{align}
Thus, by Eqs.~\eqref{eq:full-multiplet-def} and \eqref{eq:multiplet_prop_expansion}, the physical $\varphi$ state's deviation from a non-interacting state is perturbatively expanded as
\begin{align}
    \delta u_\varphi = \mathcal G \mathcal M u_\varphi = (\mathcal{G}_0 + \mathcal{G}_0 \mathcal M \mathcal{G}_0 + \cdots) \mathcal M u_\varphi = \mathcal{G}_0 \mathcal M u_\varphi + \mathcal O\left(\left(\kappa/\lambda\right)^2\right).
\end{align}
This gives the physical polarizations in the fully interacting theory, at leading order in weak mixing, as 
\begin{align}
\label{eq:physical_polarizations}
    \widetilde u_h^I(p_h)= \begin{pmatrix}
    \varepsilon\\[2mm] 
    -\mathcal K_{\pi\pi}^{-1}(p_h) \mathcal K_{\pi h}(p_h)\varepsilon
    \end{pmatrix} + \mathcal O\left(\left(\kappa/\lambda\right)^2\right),
    \hskip 1 cm
    \widetilde u_\pi^I(p_\pi)= \begin{pmatrix}
    -\mathcal K_{hh}^{-1}(p_\pi) \mathcal K_{h \pi}(p_\pi)e\\[2mm] 
    e
    \end{pmatrix} + \mathcal O\left(\left(\kappa/\lambda\right)^2\right),
\end{align}
where $\mathcal K_{\pi\pi}^{-1}(p_h)$ and $\mathcal K_{hh}^{-1}(p_\pi)$ are regular, due to the on-shell conditions $p_h^2=0$ and $p_\pi^2 = (c_s^2 -1)|\mathbf p_\pi|^2 \neq 0$. Notably, pole shifts do not enter until second order in $\kappa/\lambda$ (see, e.g., \eqn{eq:jeans-shift}), so they can be neglected in our leading order analysis.

An important point is that if we use the TT gauge for the external graviton polarization $u_h$, then the physical graviton state $\widetilde u_h$ satisfying the full non-diagonal on-shell condition Eq.~\eqref{eq:full-onshell-condition} is exactly the TT graviton polarization of pure gravity.  Using Eq.~\eqref{eq:nonint_onshell_conditions}, the full interaction matrix acting on a decoupled graviton state yields a term contracting the graviton polarization with the kinetic mixing term,
\begin{align}
    \mathcal K_{IJ} \begin{pmatrix}
    \varepsilon_{(i)}^{\mathrm{TT}}\\[2mm] 
    0
    \end{pmatrix} = 
    \begin{pmatrix}
        \mathcal K_{hh} \varepsilon^\text{TT}_{(i)}\\[2mm]
        \mathcal K_{\pi h} \varepsilon^\text{TT}_{(i)}
    \end{pmatrix}
    =
    \frac{\kappa}{2\lambda}\begin{pmatrix}
    0\\[2mm] 
    {\mathcal B}_a{}^{\rho\sigma}\varepsilon_{\rho\sigma,(i)}^{\mathrm{TT}}
    \end{pmatrix}.
\end{align}
However, this term vanishes for a TT graviton.
\begin{align}
\label{eq:no_ext_grav_to_phon}
{\mathcal B}_a{}^{\rho\sigma}\varepsilon_{\rho\sigma,(i)}^{\mathrm{TT}} &=
\left[\left(\eta^{\rho0}\eta^{\sigma0} + c_s^2\sum_{b=1}^{D-1}\eta^{\rho b}\eta^{\sigma b} \right) \partial_a + 2 \sum_{c=1}^{D-1}\eta^{0(\rho} \eta^{\sigma)c} \delta_{ca}\,\partial_0\right]\varepsilon_{\rho\sigma,(i)}^{\mathrm{TT}}\notag\\ 
&= \partial_a\varepsilon_{00,(i)}^{\mathrm{TT}} 
+ c_s^2\sum_{b=1}^{D-1} \partial_a \varepsilon_{bb,(i)}^{\mathrm{TT}} 
- \partial_0 \left(\varepsilon_{0a,(i)}^{\mathrm{TT}} + \varepsilon_{a0,(i)}^{\mathrm{TT}} \right) \notag \\
&=0 \,,
\end{align}
where the last equality uses
$\varepsilon_{0\nu,(i)}^{\rm TT}=0$ and
$\sum_{b=1}^{D-1}\varepsilon_{bb, (i)}^{\rm TT}=0$, as in Eqs.~\eqref{eq:TT-conditions} and \eqref{eq:fix-residual-gauge-freedom}.
Thus, a TT graviton satisfies the full non-diagonal on-shell condition Eq.~\eqref{eq:full-onshell-condition} in the interacting theory,
\begin{align}
    \left(\widetilde u_{h,(i)}^\text{TT}\right)^I= \begin{pmatrix}
    \varepsilon_{\mu\nu,(i)}^{\mathrm{TT}}\\[2mm] 
    0
    \end{pmatrix}.
\end{align}
This statement is exact to all orders in insertions of the kinetic mixing within the quadratic action of \autoref{sec:quadratic_terms}. The $\mathcal O(\kappa^2w_0h^2)$ terms dropped there act on the TT modes but are suppressed by $r^2/\ell_G^2$.

Fixing the residual gauge freedom Eq.~\eqref{eq:fix-residual-gauge-freedom} picks the representative physical polarization where the gauge term is kept only in the graviton external wavefunction at leading order in the weak mixing parameter $\kappa / \lambda$.   As a result, any physical TT graviton cannot kinetically mix into a longitudinal phonon. This is not generally manifest for other choices of external polarization. These TT polarizations show transparently the gauge-invariant symmetry selection rule that, at fixed graviton momentum $\mathbf q_i$, a homogeneous isotropic background preserves $\mathrm{SO}(D-2)$, under which the TT tensor and longitudinal scalar belong to inequivalent representations.  This selection rule will be useful in the fluid rest frame to simplify the list of diagrams that must be considered for a physical process, as discussed in \autoref{sec:ScatteringAmplitudes}.

\subsection{Feynman rules}
\label{sec:feynman_rules}

Here we collect the propagators and the three-point Feynman rules in the non-interacting basis. We summarize the conventions for the Feynman rules and the operators used to write them compactly in \app{app:conventions}. 
One must be careful about the signs of the metric when writing Feynman rules that explicitly mix Lorentz- and Cartesian-index objects.  Phonons carry Cartesian Euclidean indices, so phonon-space operators may be written with either upper or lower spatial indices without introducing a sign.  This remains true for a phonon-space index on a mixed object, whereas converting a Lorentz index to a spatial index can introduce a sign.  We take the convention that spatial indices in the Feynman rules are Cartesian Euclidean indices and are always written upstairs, while Lorentzian conversion signs are written explicitly or carried by factors such as \(\eta^{\mu a}\), as appropriate. Fortunately, the on-shell Ward identities in \autoref{sec:GenWardId} provide independent checks of these relative signs in the Feynman rules.

\begin{figure}[tb]
\centering
\ManuscriptFigurePDF[scale=1]{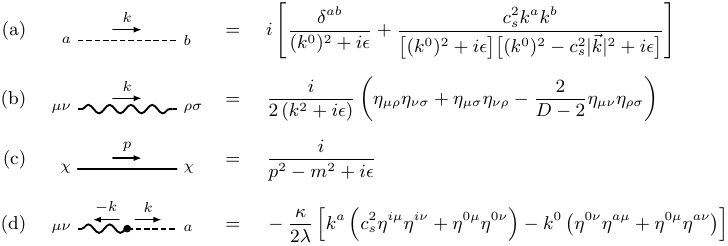}%
\caption{(a) Phonon, (b) graviton and (c) massive scalar probe propagators in the non-interacting basis.
The kinetic mixing (d) represented by a two-point vertex $\bullet$ converts a graviton
into a phonon, and denotes an explicit
$h\pi$ mixing insertion of order $\kappa/\lambda$ when treating the off-diagonal terms perturbatively.
Repeated spatial indices are implicitly summed: $A^iB^i \equiv \sum_{i=1}^{D-1}A^iB^i$.} 
\label{fig:propagators}
\end{figure}

\begin{figure}[p]
\centering
\ManuscriptFigurePDF[scale=1]{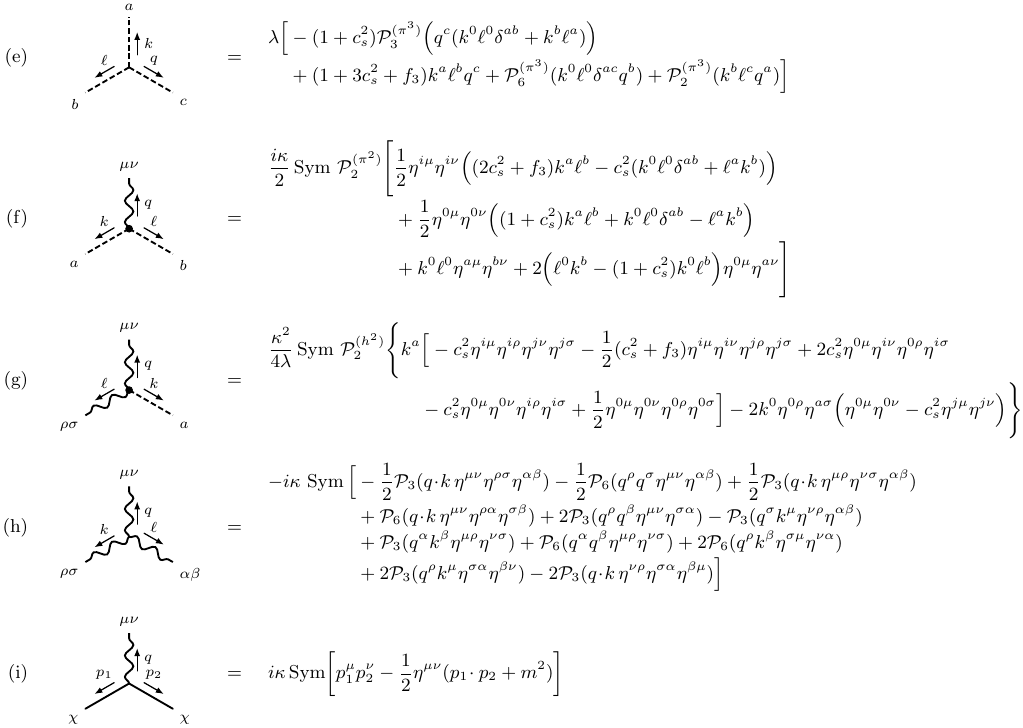}
\caption{Three-point vertices: (e) $\pi^3$ flat-space phonon interaction, (f) $h\pi^2$ interaction, (g) $h^2\pi$ interaction, (h) $h^3$ pure gravity interaction, and (i)
$h\chi^2$ gravity-probe interaction.
Filled circles $\bullet$ denote vertices with both phonons and gravitons, while vertices that are pure-phonon, pure-gravity, or gravity-probe interactions do not have an explicit vertex dot.}
\label{fig:eq33-three-point-vertices}
\end{figure}

The non-interacting basis propagators and the graviton-phonon two-point vertex are collected in Fig.~\ref{fig:propagators}, while the three-point interaction vertices are collected in Fig.~\ref{fig:eq33-three-point-vertices}. The four-point vertices are given in \app{app:four-point-ward-identities}.  We use them here to explicitly verify the Ward identities in a nontrivial case.

\clearpage

These Feynman rules are ultimately intended for higher-order calculations where ultraviolet and infrared divergences are encountered. In gravity, dimensional regularization is typically the regulator of choice because it preserves gauge invariance and meshes well with existing multi-loop integration techniques, so we explicitly write the Feynman rules in $D$ dimensions. For theories such as the fluid Lagrangian \eqref{eq:GeneralLagrangian}, which contains the Levi-Civita symbol, analytic continuation away from $D=4$ can involve additional subtleties. These issues have been studied extensively in chiral gauge theories, Chern-Simons theory, and self-dual field theories (see, for example, Refs.~\cite{tHooft:1972tcz, Chen:1992ee, Giavarini:1992ww, Chaichian:1998tf, Chetyrkin:1998mw,  Bruque:2018bmy}). The Ward identities discussed in \autoref{sec:GenWardId} provide consistency conditions to resolve any encountered subtleties.

Effects outside the minimal rules, such as accretion, finite-size response, or shifts in black-hole spectra, should enter through matched operators rather than changes to the universal pure-gravity vertices \cite{Barausse:2014tra, Pezzella:2024tkf}.  We therefore use the rules above as the minimal building blocks for the leading environmental observables and for higher-order corrections.

\section{Scattering amplitudes}
\label{sec:ScatteringAmplitudes}
\begin{figure}[tb]
\centering
\ManuscriptFigurePDF[scale=1]{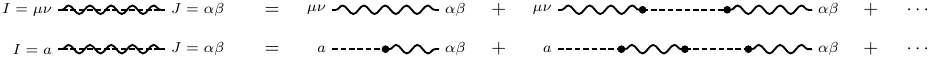}
\caption{Some representative examples of the multiplet's exact two-point Green's function, expanded in the non-interacting basis. The multiplet field is represented diagrammatically with a superimposed wavy-dashed line, while the non-interacting basis graviton and phonon fields are represented as wavy and dashed lines, with filled circles $\bullet$ denoting kinetic mixing insertions of order $\kappa / \lambda$.}
\label{fig:multiplet-exact-two-point-green-function}
\end{figure}

To identify the scattering amplitudes relevant for a given process at leading order in weak mixing, one must write down all allowed topologies in the multiplet basis \eqref{eq:GravitonPhononMultiplet} at the desired order in perturbation theory, then expand perturbatively in the weak mixing parameter $\kappa/\lambda$.  Some heuristics for writing down these amplitudes in the fluid rest frame's non-interacting basis can be drawn from the discussion in \autoref{sec:physical_state_conditions}:

For internal legs in a multiplet basis diagram, the propagator matrix expands simply in the non-interacting basis, as shown in \eqn{eq:multiplet_prop_expansion} and diagrammatically in \autoref{fig:multiplet-exact-two-point-green-function}. The full-multiplet propagator is represented as a superimposed wavy-dashed line, reflecting that the graviton and phonon degrees of freedom in the fully interacting theory are tied together.  Though at leading order in $\kappa / \lambda$, the multiplet propagator from graviton-to-graviton or phonon-to-phonon indices is just the respective non-interacting graviton or phonon propagator.  Similarly, when the multiplet propagates from graviton indices to phonon indices, or vice versa, it is a single kinetic mixing insertion of the non-interacting basis graviton and phonon at leading order.  Thus one need only consider internal legs that are non-interacting propagators or have at most one kinetic mixing insertion.

\begin{figure}[b]
\centering
\ManuscriptFigurePDF[scale=0.89]{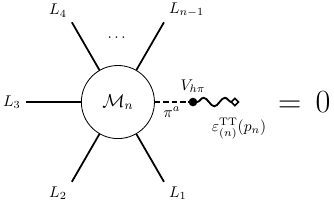}
\vspace{-0.25cm}
\caption{Selection rule that diagrams involving phonon-to-graviton conversions on external legs vanish upon contraction with an on-shell physical TT graviton polarization tensor. Each solid external line $L_i$ is a generic placeholder and may represent a multiplet, matter leg, phonon, or graviton leg, either on shell or off shell.
The filled circle $\bullet$ labeled $V_{h\pi}$ denotes the explicit $h\pi$ conversion vertex, which is of order $\kappa/\lambda$, while the diamond $\diamond$ marks the physical on-shell state, which is a TT-graviton here.}
\label{fig:n-Point_TT_ConvertVanish}
\end{figure}

For physical external states, we found above that at leading order in weak mixing, the poles are unshifted, and the physical polarizations \eqn{eq:physical_polarizations} are nearly the original field polarizations.  If we use the TT gauge for the external graviton polarizations, then the diagrams with an external on-shell graviton mixing into a phonon all vanish by \eqn{eq:no_ext_grav_to_phon}, and they can immediately be dropped.  This selection rule is illustrated in \autoref{fig:n-Point_TT_ConvertVanish}.  As a result, external legs of scattering amplitudes in the non-interacting basis can either be an on-shell phonon with no mixing, an on-shell phonon that mixes exactly once into a graviton, or an on-shell TT graviton.

Since the kinetic mixing insertion is suppressed by the weak mixing parameter $\kappa/\lambda$, no legs have more than one mixing insertion, which is apparent in \autoref{fig:multiplet-exact-two-point-green-function}.  As a result, vacuum-gravity calculations in the presence of a weakly self-gravitating fluid, like the Newtonian potential between two massive objects, are not modified at leading order in weak mixing.  One could even take the perspective that this preservation of pure-gravity calculations in the presence of the fluid is the natural requirement to impose when using the non-interacting basis, and then would arrive at the same conclusion as \autoref{sec:scaling} that the fluid must be weakly self-gravitating.  In general, non-interacting-basis scattering amplitudes at leading order in weak mixing can be separated into vacuum-gravity and flat-space-fluid sectors, which can be seen in the diagrams of the leading four-point on-shell Ward identity in \autoref{fig:4-pt-diagonal-ward}.  

\autoref{tab:tree-level-phonon-emission-amplitudes} reports the tree-level amplitudes for on-shell one- and two-phonon emission in integer dimension $D$. As demonstrated in \autoref{sec:DF_from_Amplitudes}, the single-phonon emission amplitude in \autoref{tab:tree-level-phonon-emission-amplitudes}(a) contains all the information needed to compute the leading-order relativistic dynamical friction force for linear motion.

\begin{table}[b]
\centering
\ManuscriptFigurePDF[scale=0.94]{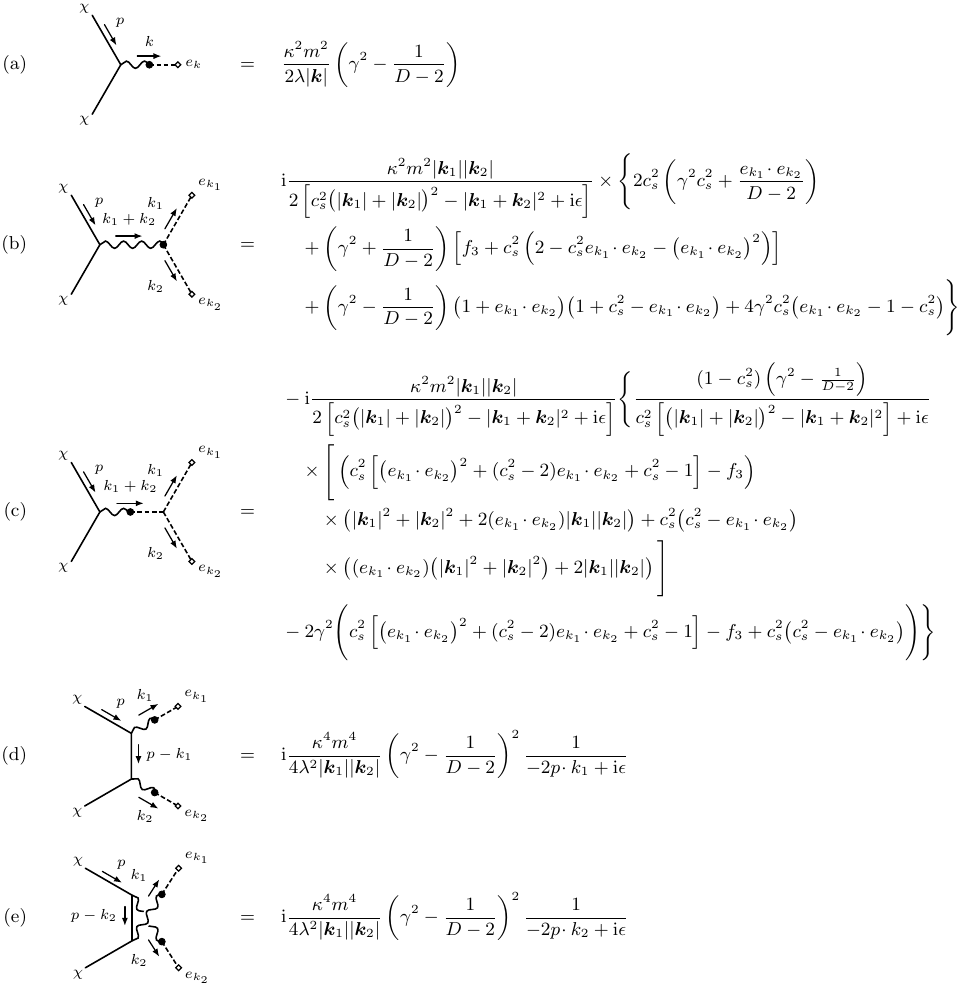}
\vspace{-0.25cm}
\caption{Tree-level one- and two-phonon emission
amplitudes from a massive perturber in integer dimension $D$ under the kinematic assumption $p\cdot k_i = 0$, expected in the eikonal limit. Solid, wavy, and dashed lines denote the perturber, graviton, and phonon, respectively. Filled circles $\bullet$ denote interactions with both phonons and gravitons, including kinetic mixing insertions. Diamonds $\diamond$ mark on-shell external states in the non-interacting basis, which here are longitudinal phonons with polarization $e_\ell$ with momentum $\ell$, and the external $\chi$ legs are implicitly on the mass shell, $p^2 = m^2$.}
\label{tab:tree-level-phonon-emission-amplitudes}
\end{table}
\clearpage

\section{Generalized on-shell Ward identities}
\label{sec:GenWardId}

\subsection{Structure of the on-shell Ward identities}

\begin{figure}[tb]
\ManuscriptFigurePDF[scale=1]{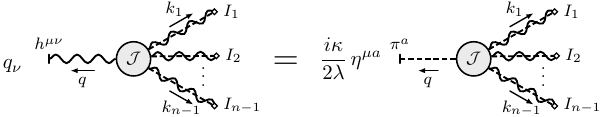}
\caption{General structure of on-shell Ward identities in the full graviton-phonon multiplet basis.
The amputated graviton leg with outgoing momentum $q$ connected to a conserved current $\mathcal J$ contracts with its momentum and becomes the same graph with a phonon, up to a projection of indices. The bar at the end of the amputated graviton and phonon legs denotes open roots with no conditions imposed on them. Diamonds $\diamond$ mark the external on-shell spectator multiplet legs with outgoing momenta $k_i$, and the labels $I_i$ specify the polarization.
The same spectators are held fixed on both sides of the identity.}
\label{fig:GeneralWardIdentity}
\end{figure}

On-shell Ward identities provide a powerful way to verify Feynman rules and that the resulting scattering amplitudes have been computed correctly,
especially for higher-order processes with infrared and ultraviolet divergences which require regularization.
This is particularly important in theories, such as the one in this paper, where the Levi-Civita symbol appears in the action (see \autoref{sec:feynman_rules} for further discussion). In addition, fluid EFTs can have infrared subtleties associated with the transverse phonon modes~\cite{Endlich:2010hf, Goldberger:2025mgb}.
A key consistency requirement for dealing with any such subtleties is that the Ward identities be satisfied.  At their basic level, the Ward identities are statements that gauge bosons and gravitons must couple to conserved currents, where the currents are defined in terms of the on-shell states of the system. 

Consider an $n$-point scattering amplitude in an unbroken gravitational system,
\begin{equation}
\mathcal M_n = \pol_{\mu\nu} \mathcal M_{h, n}^{\mu \nu}(q, u_1, u_2, \cdots, u_{n-1}) \,, 
\end{equation}
where we have exposed the polarization tensor of the graviton carrying momentum $q$.  The $u_i$ signify any particles satisfying on-shell conditions.  
The basic gravitational on-shell Ward identity requires that under the replacement $\pol_{\mu\nu}(q) \rightarrow  q_\nu \xi_\mu$, with  $\xi \cdot q = 0$, the amplitudes must vanish~\cite{Weinberg:1964ew}, 
\begin{equation}
q_\nu \xi_\mu\mathcal M_{h, n}^{\mu \nu}(q, u_1, u_2, \cdots, u_{n-1}) = 0 \,, \hskip 2 cm  \xi \cdot q = 0\,.
\label{eq:WardUnbroken}
\end{equation}
The condition on $\xi$ eliminates any terms where $\mathcal M_{h,n}^{\mu\nu}$ is shifted by a nonvanishing function proportional to $q^\mu$, which has no effect on the scattering amplitudes.
This identity is remarkably simple, as it is a statement of ordinary current conservation and is not corrected by nonlinear terms as occurs with covariant conservation.

For our case, the gauge invariance \eqref{eq:multiplet-gauge-transformation-position} mixes gravitons and phonons.  This modifies the $n$-point on-shell Ward identity to 
\begin{equation}
 q_\nu\mathcal M_{h,n}^{\mu\nu}
 (q;\widetilde u_1,\ldots,\widetilde u_{n-1})
 =\frac{i\kappa}{2\lambda}\,\eta^{\mu a}
 \mathcal M_{\pi^a,n}
 (q;\widetilde u_1,\ldots,\widetilde u_{n-1}) .
\label{eq:full-mixed-ward}
\end{equation}
where, in general, the free $\mu$ index should be taken as contracted into a transverse $\xi^\mu$ as in \eqn{eq:WardUnbroken}, though the identities can hold even without this condition, as they do for the three-point and four-point tree-level examples in this paper.   The spectator legs in this identity are on-shell states in the full multiplet basis, as shown diagrammatically in \autoref{fig:GeneralWardIdentity}.  The multiplet form of the on-shell Ward identity \eqn{eq:full-mixed-ward} is a natural starting point for writing down the leading Ward identities in $\kappa/\lambda$ as we do below. 

This identity is similar to that for a spontaneously broken gauge theory, where gauge invariance mixes with the would-be Goldstone boson, resulting in a modified on-shell Ward identity. Discussions of the need for these generalized Ward identities in simple examples of broken gauge theories such as superconductivity go back to the 1960s \cite{Nambu:1960tm}, and they have been studied more generally in modern literature on ``Goldstone Equivalence'' (see, e.g., Refs.~\cite{Cuomo:2019siu, Kribs:2022gri}). 

In our generalized Ward identity, the graviton leg contracted with its momentum, which is like an unphysical longitudinal polarization in an unbroken gauge theory, is now augmented with the same graph where the graviton is replaced by a phonon leg.  The homogeneous fluid background selects a preferred rest frame and spontaneously breaks boost invariance. 
The presence of the fluid therefore breaks spatial diffeomorphisms, producing a physical longitudinal graviton mode that propagates subluminally \cite{Lin:2015cqa, Aoki:2022ipw}, which is transparent in the unitary gauge discussed in \autoref{sec:quadratic_terms}. As a result, the generalized on-shell Ward identity in \autoref{fig:GeneralWardIdentity} illustrates the fact that we can choose a basis in which the subluminal longitudinal graviton mode can be captured entirely by the phonon degree of freedom. This is a gravitational analog of the phenomenon observed in electromagnetic waves propagating in a charged superfluid (superconductor), in which the photon picks up a physical longitudinal mode that propagates subluminally \cite{Gabadadze:2007si, Gabadadze:2008pj, Nieves:2024viy}.

\subsection{Three-point Ward identities}

\begin{figure}[!t]
    \centering
    \ManuscriptFigurePDF[scale=1]{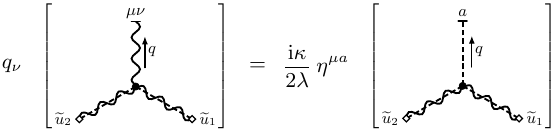}
    \caption{Three-point on-shell Ward identity in the full graviton-phonon multiplet basis.  The open root is contracted only with its outgoing momentum $q_\nu$, since the $\mu$ index need not be contracted with a polarization at tree level.  Each external polarization $\widetilde u_i$ is a physical solution of the full non-diagonal on-shell equation \eqref{eq:full-onshell-condition}, 
    $\mathcal K_{IJ}(p_i,-p_i)\widetilde u_i^J=0$.}
    \label{fig:3-pt-multiplet-ward}
\end{figure}

We follow the discussion in \autoref{sec:ScatteringAmplitudes} to expand in the weak mixing parameter $\kappa/\lambda$ and obtain perturbative on-shell Ward identities in the non-interacting basis. Below we keep only the leading terms in $\kappa/\lambda$ when applying the on-shell Ward identities, which are useful for verifying the Feynman rules needed for nonlinear fluid interactions in post-Minkowskian environmental corrections.  As long as the hierarchies in \autoref{sec:scaling} are satisfied, the Ward identities are valid, including for arbitrary Wilson coefficients that encode the equation of state.

The tree-level 3-point on-shell Ward identity at the level of the multiplet is shown in \autoref{fig:3-pt-multiplet-ward}. Its superimposed wavy-dashed line spectator legs represent the full mixed wavefunctions. The needed three-point amplitudes in the non-interacting basis at leading order in the perturbative expansion in weak mixing have two on-shell spectator legs and one open root with momentum $q$, and are given in \autoref{tab:diagonal-three-point-amplitudes}. It is worth noting that, as discussed in \autoref{sec:ScatteringAmplitudes}, there are no leading-order diagrams in the on-shell Ward identities that have a TT graviton mixing into a phonon, nor a leg with more than one mixing insertion.

\clearpage
\begin{table}[tb]
\centering
\ManuscriptFigurePDF[scale=1]{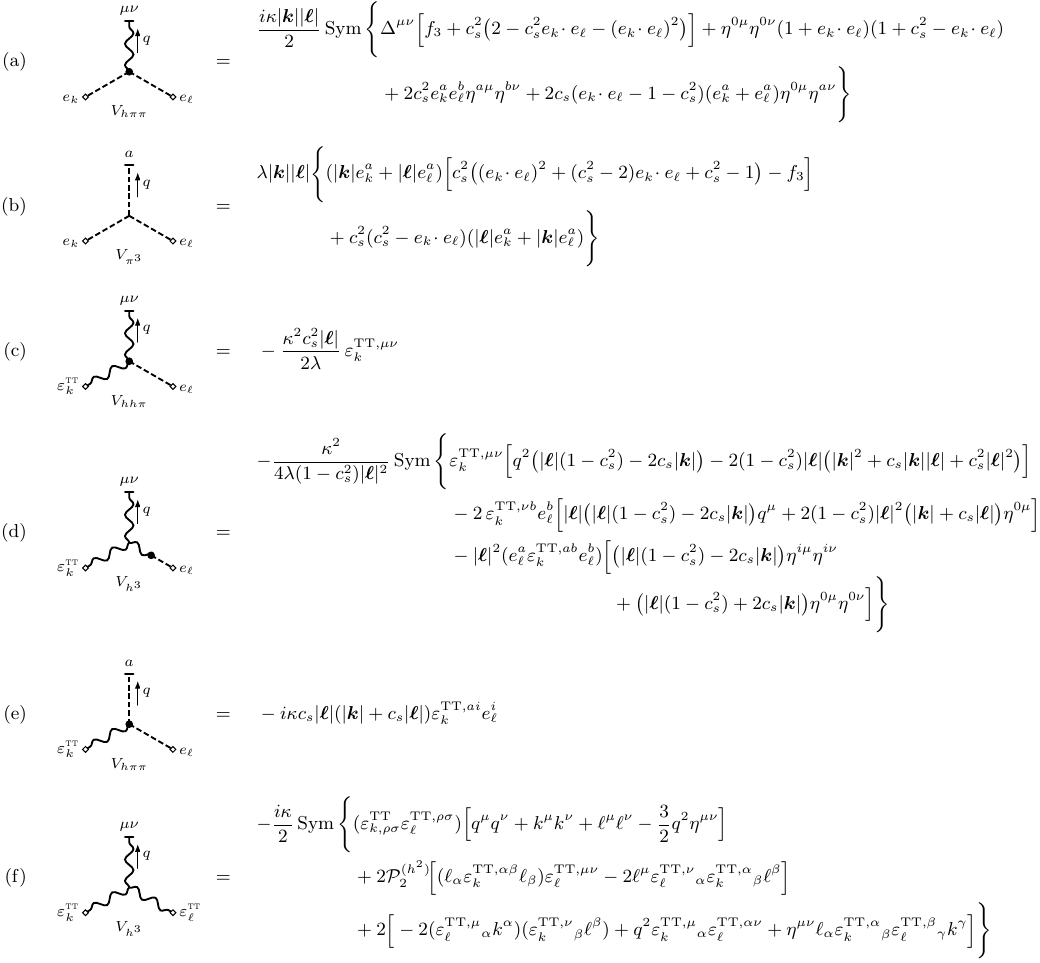}
\caption{Feynman diagrams contributing to the amputated three-point
open-root currents with two physical on-shell spectators.
The upper leg of outgoing momentum $q$ carries the free root indices, and the two lower legs of momenta $k, \ell$ are physical external states.
Rows (c) and (d) are the direct and converted-spectator contributions to the
same $h\pi$-spectator graviton-root current. Each unnormalized
$\mathcal P_2$ exchanges complete physical spectator data only, while
the root $q,\mu,\nu$ (or $q,a$) is held fixed.
Filled circles $\bullet$ represent interactions with both phonons and gravitons, bars mark amputated roots, and diamonds $\diamond$ mark
on-shell spectators in the non-interacting basis.}
\label{tab:diagonal-three-point-amplitudes}
\vspace{2.0cm}
\end{table}
\clearpage

\begin{figure}[tbp]
\centering
\ManuscriptFigurePDF[scale=1]{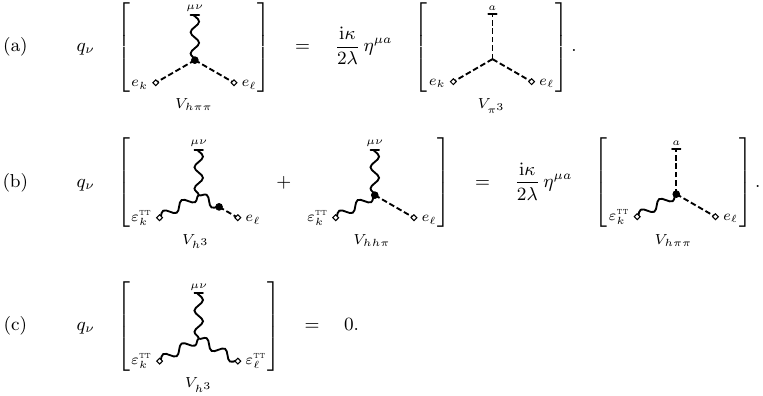}
\vspace{1.8em}
\caption{Three-point on-shell Ward
identities in the non-interacting basis at their first nonvanishing orders in $\kappa / \lambda$.
Rows (a)--(c) contain, respectively, two longitudinal-phonon, mixed
TT-graviton-longitudinal-phonon, and two TT-graviton spectator legs.
All momenta are outgoing, $q+k+\ell=0$, and the barred upper leg is the
open root carrying momentum $q$. Diamonds $\diamond$ mark physical on-shell
spectators, and the filled circles $\bullet$ denote fluid--gravity interactions.}
\label{fig:three-point-diagonal-ward-sample}
\end{figure}

After imposing the TT condition on external graviton spectator legs, the leading graph sets needed in the three spectator channels can be written as the diagrammatic equations shown in \autoref{fig:three-point-diagonal-ward-sample}. These are in terms of the open-root currents in Table~\ref*{tab:diagonal-three-point-amplitudes} and the weak-mixing physical states derived above.  In each check, $q_\nu$ contracts only the free
Lorentz index $\nu$ on the open graviton root, not acting on either
spectator leg.  Strictly speaking, contracting the other free Lorentz index $\mu$ with a physical state polarization or conserved current is generally necessary for loop-level on-shell Ward identities \cite{Bern:2014vva}, but this condition is not needed at tree level.  The open phonon-root current is instead multiplied by
$\ii\kappa\eta^{\mu a}/(2\lambda)$, as in
\eqref{eq:full-mixed-ward}.

Stepping through the different cases for the leading contributions to the Ward identity in $\kappa/\lambda $,  illustrated in \autoref{fig:three-point-diagonal-ward-sample}, we have:

\begin{itemize}
\item[(a):] \emph{Two
longitudinal-phonon spectator legs at $\mathcal O(\kappa)$.} 
Contracting the open graviton root in
Table~\ref*{tab:diagonal-three-point-amplitudes}(a) with $q_\nu$
and subtracting $\ii\kappa\eta^{\mu a}/(2\lambda)$ times the open
phonon-root current in Table~\ref*{tab:diagonal-three-point-amplitudes}(b)
reduces directly to the
diagonal phonon inverse kernels:
\begin{equation}
\label{eq:three-point-pipi-kernel-check}
\begin{aligned}
&q_\nu\mathcal M_{h,3}^{\mu\nu}(q;\pi_k,\pi_\ell)
-\frac{\ii\kappa}{2\lambda}\eta^{\mu a}
 \mathcal M_{\pi^a,3}(q;\pi_k,\pi_\ell)
\\
&\qquad =
\frac{\ii\kappa}{2}\left[
 k^\mu e_{k,b}\mathcal K_{\pi\pi}^{bc}(-\ell,\ell)e_{\ell,c}
+\ell^\mu e_{\ell,c}\mathcal K_{\pi\pi}^{bc}(k,-k)e_{k,b}
\right]
=0 \,.
\end{aligned}
\end{equation}
The terms proportional to $f_3$ cancel between the two currents, and the
remaining inverse kernels vanish by the on-shell condition
$\mathcal K_{\pi\pi}(p,-p)e_p=0$.  This establishes the $h\pi\pi$
on-shell Ward identity.

\item[(b):] \noindent\emph{One
TT-graviton, one longitudinal-phonon spectator legs at
$\mathcal O(\kappa^2/\lambda)$.}
The TT spectator has no phonon
component, whereas the physical phonon contains the graviton component
already obtained from the full on-shell condition:
\begin{equation}
\label{eq:three-point-mixed-physical-states}
\begin{aligned}
 \mathcal K_{\pi h}(k)\varepsilon_k^{\mathrm{TT}}&=0,
 &
 \widetilde u_\pi^I(\ell)&=
 \begin{pmatrix}
 -\mathcal K_{hh}^{-1}(\ell)\mathcal K_{h\pi}(\ell)e_\ell\\[1mm]
 e_\ell
 \end{pmatrix}
 +\mathcal O\!\left((\kappa/\lambda)^2\right).
\end{aligned}
\end{equation}
Thus the opposite TT-to-phonon conversion is absent, as discussed above.  The difference between
the $q_\nu$ contraction of the open graviton root in
Table~\ref*{tab:diagonal-three-point-amplitudes}(c) and
$\ii\kappa\eta^{\mu a}/(2\lambda)$ times the open phonon-root current in
Table~\ref*{tab:diagonal-three-point-amplitudes}(e) leaves the
nonzero remainder
\begin{equation}
\label{eq:three-point-mixed-direct-remainder}
q_\nu\mathcal M_{h,3}^{\mu\nu}(q;h_k,\pi_\ell)
 \!\big|_{\mathrm{III(c)}}
-\frac{\ii\kappa}{2\lambda}\eta^{\mu a}
 \mathcal M_{\pi^a,3}(q;h_k,\pi_\ell)
 \!\big|_{\mathrm{III(e)}}
=\frac{\kappa^2c_s|\bm k||\bm\ell|}{2\lambda}
 \varepsilon_k^{\mathrm{TT},\mu i}e_\ell^i .
\end{equation}
The final term is given by contracting $q_\nu$ with the free index on the open graviton root of the Einstein--Hilbert
$h^3$ vertex from Table~\ref*{tab:diagonal-three-point-amplitudes}(d).  The usual contracted-vertex identity produces inverse
graviton kernels on both spectator legs.  The terms acting on the TT
spectator vanish, while the kernel acting on the graviton component of the
phonon spectator cancels its adjacent $\mathcal K_{hh}^{-1}$ factor:
\begin{equation}
\label{eq:three-point-mixed-kernel-cancellation}
 \mathcal K_{hh}(\ell)\left[
 -\mathcal K_{hh}^{-1}(\ell)\mathcal K_{h\pi}(\ell)e_\ell
 \right]
 =-\mathcal K_{h\pi}(\ell)e_\ell .
\end{equation}
Because $\ell$ is on the sound shell,
$\ell^2=(c_s^2-1)|\bm\ell|^2$, so
$\mathcal K_{hh}^{-1}(\ell)$ is regular for nonzero $\bm\ell$ and
$c_s^2\neq1$.  The resulting $q_\nu$ contraction of the current in
Table~\ref*{tab:diagonal-three-point-amplitudes}(d) is
\begin{equation}
\label{eq:three-point-mixed-converted-remainder}
 q_\nu\left.
 \mathcal M_{h,3}^{\mu\nu}(q;h_k,\pi_\ell)
 \right|_{\mathrm{III(d)}}
 =
-\frac{\kappa^2c_s|\bm k||\bm\ell|}{2\lambda}
 \varepsilon_k^{\mathrm{TT},\mu i}e_\ell^i ,
\end{equation}
which cancels the remainder in
\eqn{eq:three-point-mixed-direct-remainder}. This establishes the
$hh\pi$ on-shell Ward identity and provides a sensitive check of the
relative sign of the two-point mixing insertion.

\item[(c):] \emph{Two TT-graviton spectator legs at $\mathcal O(\kappa)$.}
At this order, only the Einstein--Hilbert $h^3$ current in
Table~\ref*{tab:diagonal-three-point-amplitudes}(f) contributes.
Contracting the free index $\nu$ on its open graviton root with $q_\nu$
gives an expression entirely in terms of the diagonal graviton inverse
kernel:
\begin{equation}
\label{eq:three-point-hh-kernel-check}
\begin{aligned}
&q_\nu\mathcal M_{h,3}^{\mu\nu}(q;h_k,h_\ell)
=\frac{\ii\kappa}{2}
 \varepsilon_{k,\rho\sigma}^{\mathrm{TT}}
 \varepsilon_{\ell,\alpha\beta}^{\mathrm{TT}}
 \Big[
 k^\mu\mathcal K_{hh}^{\rho\sigma,\alpha\beta}(-\ell,\ell)
 +2\eta^{\mu(\rho}q_\tau
   \mathcal K_{hh}^{\sigma)\tau,\alpha\beta}(-\ell,\ell)
 \\[-1mm]
&\hspace{5.2cm}
 +\ell^\mu\mathcal K_{hh}^{\rho\sigma,\alpha\beta}(k,-k)
 +2\eta^{\mu(\alpha}q_\tau
   \mathcal K_{hh}^{\rho\sigma,\beta)\tau}(k,-k)
 \Big]
 =0 \,.
\end{aligned}
\end{equation}
The first two structures vanish when
$\mathcal K_{hh}(-\ell,\ell)$ acts on
$\varepsilon_\ell^{\mathrm{TT}}$, and the last two vanish when
$\mathcal K_{hh}(k,-k)$ acts on $\varepsilon_k^{\mathrm{TT}}$.
The phonon-root replacement starts only at
$\mathcal O(\kappa^3/\lambda^2)$, so it is absent from this leading check.
This establishes the $h^3$ on-shell Ward identity, which at leading order in $\kappa/\lambda$ is the same as the on-shell Ward identity in pure gravity.

\end{itemize}

The four-point on-shell Ward identities, which are of course more complicated and involve both exchange and contact diagrams, are derived and checked in \app{app:four-point-ward-identities}.

\section{Dynamical friction from phonon emission}
\label{sec:DF_from_Amplitudes}

As a basic test of the formalism, we demonstrate in this section how to compute classical gravitational observables with these Feynman rules, recovering the leading-order dynamical friction force from a tree-level amplitude for phonon emission.
A natural amplitude-based framework for computing post-Minkowskian environment effects is the formalism of Kosower, Maybee, and O'Connell (KMOC), which defines classical scattering observables as quantum expectation values and expresses their classical limits in terms of on-shell scattering amplitudes~\cite{Kosower:2018adc}. 
This formalism has proven to be very useful for vacuum post-Minkowskian calculations such as impulse, radiation, and energy loss (see, e.g., Ref.~\cite{Herrmann:2021lqe}).  
In the adiabatic limit, the dynamical-friction force can be inferred from the rate at which the massive perturber transfers energy and momentum to fluid excitations associated with the wake. For this case of a phonon excitation carrying away energy, the result is an observable-weighted unitarity cut.

Let $\widehat K_\pi^\mu$ denote the total four-momentum operator of asymptotic phonon excitations.  In the KMOC formalism, the radiated four-momentum into the fluid from a single phonon is 
\begin{equation}
R_\pi^\mu = \langle\psi_{\rm in}| T^\dagger \widehat K_\pi^\mu T |\psi_{\rm in}\rangle \,.
\end{equation}
Upon inserting a complete set of asymptotic states, this expression becomes a sum of products of an amplitude and its complex conjugate, with every intermediate state placed on shell and weighted by the four-momentum carried by its phonons. 

At leading order in the weak-mixing expansion, the only relevant final state consists of the recoiling massive perturber and one longitudinal phonon,
\begin{equation}
\chi(p)\longrightarrow\chi(p')+\pi(k) \, ,
\end{equation}
as explained below.  For canonically normalized states, the corresponding differential rate in four spacetime dimensions is
\begin{equation}\label{eq:diffGamma}
\mathrm d\Gamma(\chi \to \chi + \pi) 
=
\frac{1}{2E_p}\,
\frac{\mathrm d^3\bm p'}{(2\pi)^3\,2E_{p'}}\,
\frac{\mathrm d^3\bm k}{(2\pi)^3\,2\omega_k}\,
(2\pi)^4\delta^{(4)}(p-p'-k)\,
\left|
\mathcal A_{\rm tree}(p\rightarrow p',k)
\right|^2  .
\end{equation}
The radiated four-momentum per unit coordinate time (in the fluid rest frame) is consequently
\begin{equation}
\frac{\mathrm dP_\pi^\mu}{\mathrm dt} =
\int k^\mu\,\mathrm d\Gamma(\chi \to \chi + \pi) \,,
\end{equation}
from which we have the energy transferred to the fluid as
\begin{equation}
\frac{\mathrm dE_\pi}{\mathrm dt} = \int\omega_k\,\mathrm d\Gamma(\chi \to \chi + \pi) \,.
\label{eq:PhononEnergyRate}
\end{equation}
For rectilinear motion, the drag force is parallel and opposite to the velocity of the perturber.  Energy conservation then gives the dynamical friction,
\begin{align}
|F_\text{DF}|
& = 
\frac{1}{v}\frac{\mathrm dE_\pi}{\mathrm dt}
= \frac{1}{v} \int \omega_k \, \mathrm{d}\Gamma(\chi \to \chi + \pi)\,,
\label{eq:Force}
\end{align}
where $\omega_k$ is the energy of the on-shell phonon with momentum $k^\mu$, and the integral is taken over the allowed fluid excitation modes $k$. 

\begin{figure}[t]
\centering
\ManuscriptFigurePDF[scale=1]{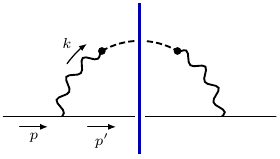}
\caption{Leading order in $\kappa/\lambda$ energy-weighted unitarity cut (vertical blue line) for single-phonon emission mediated by gravity, which gives the leading order dynamical friction force for supersonic motion in an inviscid fluid.  
Filled circles $\bullet$ denote the two explicit $h\pi$ mixing
insertions. The solid line denotes a
compact-object worldline, wavy lines denote gravitons, and dashed lines denote phonons. The self-energy loop carries momentum $k$, and legs cut by the vertical blue line are physical on-shell states in the non-interacting basis.}
\label{fig:df-leading-kmoc-cut}
\end{figure}

At the leading order in the weak-mixing expansion, the dynamical friction force has only one nonzero contribution to this energy-weighted unitarity cut in the non-interacting basis, shown diagrammatically in \autoref{fig:df-leading-kmoc-cut}.  This picture can be interpreted as the leading-order dynamical friction force resulting from a single-phonon emission, in which the compact object gravitationally excites a wake in the fluid (on-shell longitudinal phonon), and this wake gravitationally pulls back on the object.  In particular, the graviton cuts of this self-energy diagram vanish due to the selection rule condition shown in \autoref{fig:n-Point_TT_ConvertVanish} that an on-shell TT graviton does not mix with a phonon; the on-shell transverse phonon has zero energy ($\omega_k^2=0$), so it does not contribute to energy-weighted observables either; and multi-phonon emission is higher order and not considered for the leading calculation.

To evaluate Eq.~\eqref{eq:Force}, we work in the fluid
rest frame and take
\begin{equation}
p^\mu=(\gamma m,0,0,\gamma mv)\,, \qquad
k^\mu=(\omega_k,\bm k)\,, \qquad
\omega_k=c_sK\,, \qquad
K\equiv|\bm k|\,.
\end{equation}
The on-shell condition for the recoiling perturber,
$(p-k)^2=m^2$, fixes the emission angle to
\begin{equation}
\cos\theta^*(K)
=
\frac{c_s}{v}
+
\frac{(1-c_s^2)K}{2\gamma mv}.
\label{eq:phonon-emission-angle}
\end{equation}
 
In four spacetime dimensions, the leading differential rate $\mathrm d \Gamma(\chi \to \chi + \pi)$ reduces to an integral of the tree-level amplitude for single-phonon emission over the magnitude of the spatial momentum $K\equiv |\mathbf k|$ of the emitted on-shell phonon.
\begin{equation}
   |F_\text{DF}| = \frac{1}{16\pi\,\gamma^2\,m^2\,v^2}
\int_{K_\text{min}}^{K_\text{max}} \mathrm{d}K\; K\;
\bigl|\mathcal A_\text{tree}^{(D=4)}\!\bigl(K,\theta^*(K)\bigr)\bigr|^{2}\;
\Theta\!\Bigl(1-\bigl|\cos\theta^*(K)\bigr|\Bigr) \, .
\end{equation}
The step function $\Theta$ guarantees there is phase space for the phonon to be emitted at the physical angle $\theta^*$.
\eqn{eq:phonon-emission-angle} reduces to $\cos\theta^*= c_s/v$ in the eikonal limit, so the phonon only has phase space to be emitted for supersonic motion, $v > c_s$.  By the same argument, we again see that no graviton cuts contribute at leading order in $G$, since an emitted graviton would have on-shell phase space only for $v>1$, which is unphysical.  At tree level, the single-phonon emission amplitude in integer dimension $D$ is given in \autoref{tab:tree-level-phonon-emission-amplitudes}(a). Specializing to $D=4$,
\begin{equation}\label{eq:treelevel_amplitude}
\begin{aligned}
\mathcal A_\text{tree}
&= \frac{(2\gamma^2 - 1)\, \kappa^2 m^2}{4 \lambda |\mathbf k|}\\
&= \frac{1}{K}\,\gamma^2 (1+v^2)\,
8\pi G m^2 \sqrt{w_0} \,.
\end{aligned}
\end{equation}
This gives the leading-order relativistic dynamical friction force,
\begin{equation}\label{eq:FDF}
\begin{aligned}
\left|F_\text{DF}\right|
&= \frac{4 \pi G^2 m^2 \gamma^2 (1+v^2)^2 w_0}{v^2}\,
\log\left( \frac{K_\text{max}}{K_\text{min}} \right),
\end{aligned}
\end{equation}
for supersonic linear motion in an inviscid fluid, reproducing the coefficient of the logarithm in the steady-state result of Ref.~\cite{Barausse:2007ph}. \eqn{eq:FDF} is the magnitude of the momentum-loss rate in the fluid rest frame per unit coordinate time $t$. Note that some literature reports the momentum change in the fluid rest frame per unit proper time of the perturber, $\mathrm{d}\tau = \mathrm{d}t/\gamma$, which results in an extra Lorentz factor $\gamma$ compared to \eqn{eq:FDF}, or in the perturber rest frame per unit proper time, which matches \eqn{eq:FDF} as the two extra Lorentz factors cancel. For the benchmark of \autoref{sec:scaling}, chosen to illustrate the region of validity of the EFT, the orbit is inclined at $60^\circ$ and is inside the disk only $\simeq1.5\%$ of each period, so the orbit-averaged force is about a hundred times smaller than Eq.~\eqref{eq:FDF}.

The cutoffs are the edges of the window Eq.~\eqref{eq:window}, $K_\text{max}=1/R_\text{min}$ and $K_\text{min}=1/R_\text{max}$ (\autoref{tab:scales}), so the Coulomb logarithm measures the logarithmic width of the region in which the EFT applies. For the benchmark case of \autoref{fig:scales} it is $\log(K_\text{max}/K_\text{min})\simeq 8$. Ignoring accretion, the $\mathcal O(1)$ term inside the logarithm can be fixed by matching to the flow around the perturber at distances $r\lesssim R_\text{min}$, which the EFT does not resolve.  By power counting, the dynamical friction force at higher orders in the $G$ expansion will be power-law divergent, which requires careful treatment. Ref.~\cite{Modrekiladze:2026twz} resolves this via the formal renormalization of the dynamical friction force  together with explicit matching functions. These functions are currently known only for a pressureless fluid, for which they are obtained from the geodesic motion of the fluid particles around the perturber~\cite{Vicente:2022ivh, Traykova:2023qyv}.

Dependence on the short-distance cutoff can be avoided by considering suitable differential quantities. For example, at leading order the differential force spectrum is
\begin{equation}
 \frac{\mathrm d|F_\text{DF}|}{\mathrm dK} =  \frac{4 \pi G^2 m^2 \gamma^2 (1+v^2)^2 w_0}{v^2}\, \frac{1}{K} \,.
\label{eq:differential_force}
\end{equation}
Within the regime of validity of the EFT, this spectrum is independent of the short-distance cutoff. Consequently, the variation of the dynamical friction force under a change in the system size is a physical, cutoff-independent quantity. Since the force in \eqref{eq:FDF} and the spectrum in \eqref{eq:differential_force} share the same overall prefactor, its coefficient can be extracted from a quantity that is insensitive to the unresolved short-distance physics: the nonanalytic $1/K$ cannot be produced by any local operator. Beyond leading order, the spectrum remains cutoff independent, but its corrections are analytic in $K$ and receive contributions from contact operators whose coefficients must be fixed by matching.

At leading order, the steady-state dynamical friction force and differential force spectrum do not depend on the sound speed beyond the condition of supersonic motion, $v > c_s$, despite the Feynman rule for phonon-graviton mixing, shown in \autoref{fig:propagators}(d), used to derive the force depending on this Wilson coefficient.  There exists a field redefinition that removes the explicit $c_s^2$ from the phonon-graviton mixing, suggesting a more physical basis may exist. It would be interesting to study whether there exist other such field redefinitions that can simplify precision calculations of environmental effects.

\section{Conclusions}
\label{sec:Conclusions}

In this work, we developed an EFT framework for a relativistic compact object moving through an inviscid fluid that meshes well with the calculation tools used for the post-Minkowskian two-body problem. The construction combines the covariant effective theory of a perfect fluid with perturbative gravity and a point-particle description of the compact object, and applies when the hydrodynamic, point-particle, eikonal (classical), and weak-field expansions overlap. When the fluid background is sufficiently weakly self-gravitating, the framework simplifies further, since graviton-phonon mixing can be treated perturbatively and effects can be treated as interactions among gravitons, phonons, and the massive scalar perturber in a diagrammatic expansion.

We focused here on setting up fluid effects in the context of the post-Minkowskian expansion. To this end, we derived Feynman rules for propagators and interaction vertices with up to four fields, including the pure-fluid, mixed nonlinear graviton-phonon, pure-gravity, and gravity-probe interactions.  
We derived the corresponding on-shell Ward identities, including contributions from the inhomogeneous phonon transformation, and verified the Feynman rules through four-point order. These identities provide nontrivial checks of gauge invariance and current conservation and will be particularly valuable in dimensionally regulated higher-order calculations.

We then obtained the tree-level amplitudes for one- and two-phonon emission from the massive perturber. As a first check of our framework, the tree-level amplitude for single-phonon emission reproduces, via the KMOC formalism, the known linear-motion inviscid-fluid dynamical friction force~\cite{Barausse:2007ph}, showing that the EFT captures the wake-induced energy and momentum transfer from the compact object to the medium. More broadly, this formalism is a first step towards bridging scattering-amplitude methods for compact-binary dynamics and astrophysical environmental effects, providing control over gauge invariance, power counting, and the extraction of various observables including nonlinear effects.   

A natural next step is to extend this framework to post-Minkowskian fluid corrections in the context of binary dynamics. Such calculations will involve higher-order amplitudes, multi-phonon emission, and observable-weighted unitarity cuts~\cite{Kosower:2018adc, Herrmann:2021tct}. Higher-order contributions to dynamical friction contain short-distance power-law divergences which can be handled by formal renormalization and matching to an exact calculation of the flow past the compact object~\cite{Modrekiladze:2026twz}.  The required matching function is currently available only for collisionless matter~\cite{Vicente:2022ivh, Traykova:2023qyv}.  As noted in \autoref{sec:DF_from_Amplitudes}, dependence on the short-distance cutoff can be avoided when considering suitable differential quantities. 

Building on recent advances in amplitude methods, many important directions remain to be developed and explored, including finite-size effects through higher-dimension operators (see, e.g., Refs.~\cite{Cheung:2020sdj, Bern:2020uwk, Correia:2026utp}), compact-object spin (see, e.g., Refs.~\cite{Bern:2020buy, Jakobsen:2023ndj, Akpinar:2025bkt, Aoude:2023vdk, Bjerrum-Bohr:2026fhx}), and $N>2$-body dynamics~\cite{Loebbert:2020aos, Jones:2022aji, Solon:2024zhr}. Tidal response in a fluid background could likewise be studied using black-hole perturbation theory, perhaps by adapting recently developed techniques~\cite{Caron-Huot:2025tlq, Kosmopoulos:2025rfj, Solon:2026ubm}. Extending the present framework to generic orbital motion~\cite{Barausse:2007ph, Kim:2007zb, desjacques_analytic_2022} is nontrivial due to long-time effects, akin to complications from tail effects in vacuum gravity.  Accretion provides a complementary dissipative channel that could be incorporated through the direct transfer of mass and momentum to the compact object~\cite{Petrich:1988zz, Wong:2019yoc}. Inhomogeneous or curved backgrounds~\cite{Sanchez-Salcedo:2000gki, Vicente:2019ilr, 2026arXiv260716422G}, and intrinsically dissipative media~\cite{Crossley:2015evo, Liu:2018kfw} will also be important for applications to realistic compact binaries. More broadly, the framework presented here casts aspects of gravitational interactions with astrophysical environments in a form that can benefit from the same amplitude and integration methods that have advanced post-Minkowskian vacuum gravitational dynamics.

\begin{acknowledgments}
    We thank  Enrico Barausse, Miguel Correia, Francisco Duque, Scott Melville, Enrico Pajer, Julio Parra-Martinez, Radu Roiban, Michael Saavedra, Dam T. Son, and especially Ira Rothstein and Michael Ruf for useful discussions. We thank Enrico Barausse, Francisco Duque, Radu Roiban, and Ira Rothstein for helpful comments on this manuscript. Z.B. is supported in part by the U.S. Department of Energy (DOE) under award number DE-SC0009937. Z.B., S.D., E.H. and A.W. are supported in part by the European Research Council (ERC) Horizon Synergy Grant ``Making Sense of the Unexpected in the Gravitational-Wave Sky'' grant agreement no. GWSky-101167314. M.S. and A.W. are supported by the US Department of Energy under award number DE-SC0024224 and the Sloan Foundation. A.W. is supported by the NSF Graduate Research Fellowship under Grant No. DGE-2034835. This project used Claude, Codex, and Mathematica.  We are grateful to the Mani L. Bhaumik Institute for Theoretical Physics for support.
\end{acknowledgments}

\appendix
\section{Conventions}
\label{app:conventions}

In this appendix, we collect the conventions used in this paper. Specifically, for the Feynman rules and scattering amplitudes, our conventions are as follows:
\begin{itemize}
\item All vertices have outgoing momentum, including the phonon leg of the two-point vertex for graviton-phonon mixing in \autoref{fig:propagators}(d).
\item We use the particle theory mostly-minus metric signature, $({+}{-}\;{\cdots}\;{-})$.
\item Greek indices run from $0, \ldots, D-1$, with the flat space Levi-Civita sign convention $\tilde \epsilon^{01\cdots D-1}=+1$.  At tree level we take $D = 4$, while at loop level we often take $D= 4 - 2 \epsilon$, where $\epsilon$ is the usual dimensional regularization parameter.
\item The metric is expanded about flat space as $g_{\mu\nu} = \eta_{\mu\nu} + \kappa h_{\mu\nu}$, with
$\kappa^2 = 32 \pi G$ of canonical dimension $[\kappa]=-1$ in $D=4$. 
\item We take the eikonal limit, $k / m \ll 1$ for $k^\mu$ a phonon or graviton momentum and $m$ the mass of the perturber. This limit takes us to the classical limit.
\item Spatial indices in the Feynman rules are Cartesian Euclidean indices and are always written as upper. Repeated
spatial indices are implicitly summed, i.e. $p^i p^i =|\mathbf p|^2$. Any signs resulting from converting a Lorentz index to a spatial index are written explicitly or carried by factors such as \(\eta^{\mu a}\).
\item The dimensionless Wilson coefficients are $f'(1)=1$, $f''(1)=c_s^2$, the square of the rest fluid's sound speed, and
$f_n \equiv f^{(n)}(1)$ for $n\geq3$, which are completely fixed upon prescribing an equation of state.
\end{itemize}
The phonon fluctuation decomposes into longitudinal and transverse pieces as
\begin{equation}
\pi^i(\omega,\mathbf{k})
= \pi_\text{L}\,\hat{k}^i + \pi_\text{T}^i,
\qquad \hat{k}_i \pi_\text{T}^i = 0.
\end{equation}
The corresponding longitudinal and transverse projectors are
\begin{subequations}
\begin{align}
P_\text{L}^{ij} &= \hat{k}^i \hat{k}^j,\\
P_\text{T}^{ij} &= \delta^{ij} - \hat{k}^i \hat{k}^j.
\end{align}
\end{subequations}

The gravity sector is in De Donder gauge, a smeared gauge analogous to Feynman gauge in QED.
As discussed in \autoref{sec:physical_state_conditions}, we impose the usual transverse-traceless conditions on the graviton polarization tensor~\eqref{eq:TT-conditions} and fix the residual gauge degrees of freedom as in~\eqn{eq:fix-residual-gauge-freedom}. This physical state $\varepsilon_{\mu\nu}^\text{TT}$ is referred to throughout the text as a ``TT graviton.''

Following the typical Einstein-Hilbert vertex notation of DeWitt \cite{DeWitt:1967uc} and Sannan~\cite{Sannan:1986tz}, we introduce the operators $\operatorname{Sym}$ and $\mathcal P_n$ to encapsulate the symmetric structures of the Feynman rules and present them compactly in the text. The symmetrization operator $\operatorname{Sym}$ denotes the
normalized symmetrization of every individual graviton's indices, reflecting the symmetry of the graviton under exchange of indices $h_{\mu\nu} = h_{\nu\mu} = \frac{1}{2}h_{(\mu\nu)}$.
\begin{equation}
\operatorname{Sym}T^{\mu_1\nu_1,\ldots,\mu_r\nu_r}
\equiv \frac{1}{2^r}\prod_{s=1}^r
\bigl[1+(\mu_s\leftrightarrow\nu_s)\bigr]
T^{\mu_1\nu_1,\ldots,\mu_r\nu_r}.
\end{equation}

The permutation operator $\mathcal P_n$ is an unnormalized sum over
the $n$ permutations of the graviton or phonon leg labels that give distinct terms, reflecting the Bose symmetry of the integer-spin gravitons and phonons. For a field $\varphi \in\{h,\pi\}$ with indices set $\xi \in \{\mu\nu,a\}$, let $L^{(\varphi)}_i \equiv \varphi_{\xi_i}(p_i)$ represent the $\varphi$ leg of a diagram with the momentum-index pair labeling $(p_i,\xi_i)$. The operator $\mathcal P^{(\varphi^r)}_n$ acts on a function $T$ that depends on $r$ of these $L^{(\varphi)}$ legs, summing over all the permutations $(p_i,\xi_i) \leftrightarrow(p_j,\xi_j)$ that yield a distinct term. That is, $n = r!$ if each momentum-index pair exchange gives a new term for $T\left(L^{(\varphi)}_1,\ldots,L^{(\varphi)}_r\right)$, and $n < r!$ if any of the $L^{(\varphi)}$ swaps preserve the value of $T$. Explicitly, we can write this as
\begin{equation}
\mathcal P_n^{(\varphi^r)} T\left(L^{(\varphi)}_1,\ldots,L^{(\varphi)}_r\right)
\equiv \sum_{\sigma\,\in\, S_r/\operatorname{Stab}(T)}
T\left(L^{(\varphi)}_{\sigma(1)},\ldots,L^{(\varphi)}_{\sigma(r)}\right),
\qquad \bigl|S_r/\operatorname{Stab}(T)\bigr|=n.
\end{equation}

For example, for phonon leg labels $\{(k,a),(\ell,b),(q,c)\}$, one has in the three-phonon vertex \autoref{fig:eq33-three-point-vertices}(e)
\begin{equation}
\begin{aligned}
    \mathcal P_6^{(\pi^3)}= 
    1
    &+\Big[(k,a)\leftrightarrow(\ell,b)\Big]
    +\Big[(k,a)\leftrightarrow(q,c)\Big]
    +\Big[(\ell,b)\leftrightarrow(q,c)\Big]\\
    &+\Big[(k,a)\rightarrow(\ell,b) \rightarrow (q,c) \rightarrow (k,a)\Big]
    +\Big[(k,a)\rightarrow(q,c) \rightarrow (\ell,b) \rightarrow (k,a)\Big].
\end{aligned}
\end{equation}
As expected from the diagrammatic interpretation, these are the reflection and rotation elements of the dihedral group $D_3$. For an example with nontrivial stabilizer, consider the tensor structure
$q_1^{\mu_2}q_1^{\nu_2}\eta^{\mu_1\nu_1}
\eta^{\mu_3\nu_3}\eta^{\mu_4\nu_4}$ present in the four-graviton vertex \autoref{fig:four-point-vertices}(n), with
momentum--index pairs
$\{(q_i,\mu_i\nu_i) \,|\, i=1,\dots, 4\}$.  This structure is invariant under the exchange
$(q_3,\mu_3\nu_3)\leftrightarrow(q_4,\mu_4\nu_4)$, so its orbit contains $4!/2=12$ distinct terms,
\begingroup
\setlength{\jot}{1pt}
\begin{align}
\mathcal P_{12}^{(h^4)}\Bigl(
q_1^{\mu_2}q_1^{\nu_2}\eta^{\mu_1\nu_1}
\eta^{\mu_3\nu_3}\eta^{\mu_4\nu_4}\Bigr)
={}&q_1^{\mu_2}q_1^{\nu_2}\eta^{\mu_1\nu_1}\eta^{\mu_3\nu_3}\eta^{\mu_4\nu_4}
+q_1^{\mu_3}q_1^{\nu_3}\eta^{\mu_1\nu_1}\eta^{\mu_2\nu_2}\eta^{\mu_4\nu_4}
+q_1^{\mu_4}q_1^{\nu_4}\eta^{\mu_1\nu_1}\eta^{\mu_2\nu_2}\eta^{\mu_3\nu_3}
\notag\\[6pt]
&{}+q_2^{\mu_1}q_2^{\nu_1}\eta^{\mu_2\nu_2}\eta^{\mu_3\nu_3}\eta^{\mu_4\nu_4}
+q_2^{\mu_3}q_2^{\nu_3}\eta^{\mu_1\nu_1}\eta^{\mu_2\nu_2}\eta^{\mu_4\nu_4}
+q_2^{\mu_4}q_2^{\nu_4}\eta^{\mu_1\nu_1}\eta^{\mu_2\nu_2}\eta^{\mu_3\nu_3}
\notag\\[6pt]
&{}+q_3^{\mu_1}q_3^{\nu_1}\eta^{\mu_2\nu_2}\eta^{\mu_3\nu_3}\eta^{\mu_4\nu_4}
+q_3^{\mu_2}q_3^{\nu_2}\eta^{\mu_1\nu_1}\eta^{\mu_3\nu_3}\eta^{\mu_4\nu_4}
+q_3^{\mu_4}q_3^{\nu_4}\eta^{\mu_1\nu_1}\eta^{\mu_2\nu_2}\eta^{\mu_3\nu_3}
\notag\\[6pt]
&{}+q_4^{\mu_1}q_4^{\nu_1}\eta^{\mu_2\nu_2}\eta^{\mu_3\nu_3}\eta^{\mu_4\nu_4}
+q_4^{\mu_2}q_4^{\nu_2}\eta^{\mu_1\nu_1}\eta^{\mu_3\nu_3}\eta^{\mu_4\nu_4}
+q_4^{\mu_3}q_4^{\nu_3}\eta^{\mu_1\nu_1}\eta^{\mu_2\nu_2}\eta^{\mu_4\nu_4}.
\end{align}
\endgroup

\section{Interaction terms}
\label{app:interaction_terms}
Expanding the dimensionless function $f(b)$ about $b=1$, with Wilson coefficients
$f'(1)=1$, $f''(1)=c_s^2$, and $f_n=f^{(n)}(1)$ for $n\geq3$, the cubic action is given in \eqn{eq:three-point-action}.
The Lagrangian interaction terms in $D=4$ are 
\begingroup
\setlength{\jot}{1pt}
\begin{subequations}
\begin{align}
\mathcal L_{\pi^3}={}&\lambda
\left\{
\frac{1+c_s^2}{2}
[\partial\pi][\partial\pi^2]
-
\frac{1+3c_s^2+f_3}{6}
[\partial\pi]^3
-
\frac{1}{3}
[\partial\pi^3]
+
\frac{1+c_s^2}{2}
[\partial\pi]\dot{\pi}^{\,2}
-
\dot{\pi}^{\,i}\partial_i\pi^j\dot{\pi}^{\,j}
\right\},\\[.7em]
\mathcal L_{h\pi^2}={}&-\frac{\kappa}{4}\biggl\{
h_{00}\left((1+c_s^2)[\partial\pi]^2-[\partial\pi^2]+\dot\pi^2\right)
+h_{ii}\left((2c_s^2+f_3)[\partial\pi]^2-c_s^2([\partial\pi^2]+\dot\pi^2)\right)\notag\\
&\hskip .9 cm 
 +4h_{0i}\left(\dot\pi^j\partial_j\pi^i
-(1+c_s^2)[\partial\pi]\dot\pi^i\right)
+2h_{ij}\dot\pi^i\dot\pi^j\biggr\},\\[.7em]
\mathcal L_{h^2\pi}={}&\frac{\kappa^2}{8\lambda}\biggl\{
[\partial\pi]\Bigl(
h_{00}^2-2c_s^2h_{00}h_{ii}
-(c_s^2+f_3)h_{ii}h_{jj}
+4c_s^2h_{0i}h_{0i}-2c_s^2h_{ij}h_{ij}\Bigr)
-4(h_{00}-c_s^2h_{jj})h_{0i}\dot\pi^i
\biggr\},
\\[.7 em]
\mathcal L_{h^3}={}&-\frac{\kappa}{2}\biggl\{
\frac{h}{2}\eta^{\mu\nu}\eta^{\alpha\beta}\eta^{\rho\sigma}
-h^{\mu\nu}\eta^{\alpha\beta}\eta^{\rho\sigma}
-\eta^{\mu\nu}h^{\alpha\beta}\eta^{\rho\sigma}
-\eta^{\mu\nu}\eta^{\alpha\beta}h^{\rho\sigma}
\biggr\}\notag\\[0.3em]
&\quad\times\biggl(
\partial_\mu h_{\alpha\beta}\partial_\nu h_{\rho\sigma}
-\partial_\mu h_{\alpha\rho}\partial_\nu h_{\beta\sigma}
+2\partial_\mu h_{\beta\rho}\partial_\alpha h_{\nu\sigma}
-2\partial_\mu h_{\nu\alpha}\partial_\beta h_{\rho\sigma}
\biggr),
\\[.7 em]
\mathcal L_{h\chi^2}={}&\frac{\kappa}{2}\biggl\{\left(\frac{h}{2}\eta^{\mu\nu}-h^{\mu\nu}\right)\partial_\mu\chi\partial_\nu\chi-\frac{h}{2}m^2\chi^2\biggr\}.
\end{align}
\end{subequations}
\endgroup
Following the notation of Endlich~\cite{Endlich:2010hf}, here
$[\partial\pi]=\partial_i\pi^i$,
$[\partial\pi^2]=\partial_i\pi^j\partial_j\pi^i$, and
$[\partial\pi^3]=\partial_i\pi^j\partial_j\pi^k\partial_k\pi^i$,
while $\dot\pi^2=\dot\pi^i\dot\pi^i$.
In the pure-gravity densities, all indices are raised with the mostly-minus metric $\eta^{\mu\nu}$, and
\begin{equation*}
h\equiv\eta^{\mu\nu}h_{\mu\nu},
\qquad
(h^2)^{\mu\nu}\equiv h^\mu{}_{\lambda}h^{\lambda\nu}.
\end{equation*}

Similarly, the quartic interactions containing phonons and gravitons are
\begin{equation}
S_{}^{(4)}
=\int \mathrm{d}^Dx\left(
\mathcal L_{\pi^4}+\mathcal L_{h\pi^3}
+\mathcal L_{h^2\pi^2}+\mathcal L_{h^3\pi} +\mathcal L_{h^4} +\mathcal L_{h^2\chi^2}
\right).
\end{equation}
The six quartic interaction densities in $D=4$ are
\begingroup
\setlength{\jot}{1pt}
\begin{subequations}
\begin{align}
\mathcal L_{\pi^4}={}&\frac{\lambda^2}{24}\biggl\{
-(6f_3+f_4+7c_s^2)[\partial\pi]^4
+6(f_3+3c_s^2)[\partial\pi]^2[\partial\pi^2]
-3c_s^2[\partial\pi^2]^2-8c_s^2[\partial\pi][\partial\pi^3]\notag\\
&\quad+6(1+f_3+3c_s^2)[\partial\pi]^2\dot\pi^2
-6(1+c_s^2)[\partial\pi^2]\dot\pi^2
+3(1-c_s^2)(\dot\pi^2)^2
-24(1+c_s^2)[\partial\pi]\dot\pi^i\partial_i\pi^j\dot\pi^j\notag\\
&\quad+12\left(\dot\pi^i\partial_i\pi^j\right)
\left(\dot\pi^k\partial_k\pi^j\right)
+24\dot\pi^i\partial_i\pi^j\partial_j\pi^k\dot\pi^k
\biggr\},\\[.7 em]
\mathcal L_{h\pi^3}={}&\frac{\kappa\lambda}{12}\biggl\{
h_{00}\Bigl(
-(1+f_3+3c_s^2)[\partial\pi]^3
+3(1+c_s^2)[\partial\pi][\partial\pi^2]
-2[\partial\pi^3]
+6\dot\pi^i\partial_i\pi^j\dot\pi^j
-3(1+c_s^2)[\partial\pi]\dot\pi^2\Bigr)\notag\\
&\quad+h_{ii}\Bigl(
-(5f_3+f_4+4c_s^2)[\partial\pi]^3
+3(f_3+2c_s^2)[\partial\pi]
\left([\partial\pi^2]+\dot\pi^2\right)
-2c_s^2[\partial\pi^3]
-6c_s^2\dot\pi^i\partial_i\pi^j\dot\pi^j\Bigr)\notag\\
&\quad+6h_{0i}\dot\pi^i\Bigl(
(1+f_3+3c_s^2)[\partial\pi]^2
-(1+c_s^2)[\partial\pi^2]
+(1-c_s^2)\dot\pi^2\Bigr)
-12(1+c_s^2)[\partial\pi]h_{0i}\dot\pi^j\partial_j\pi^i\notag\\
&\quad-6(1+c_s^2)[\partial\pi]h_{ij}\dot\pi^i\dot\pi^j
+12h_{0i}\dot\pi^j\partial_j\pi^k\partial_k\pi^i
+12h_{ij}\dot\pi^i\dot\pi^k\partial_k\pi^j
\biggr\},\\[.7 em]
\mathcal L_{h^2\pi^2}={}&\frac{\kappa^2}{16}\biggl\{
[\partial\pi]^2\Bigl(
(1+c_s^2)h_{00}^2
+4(f_3+2c_s^2)h_{0i}h_{0i}
-2(f_3+2c_s^2)h_{ij}h_{ij}
-2(f_3+2c_s^2)h_{00}h_{ii}
-(4f_3+f_4+2c_s^2)h_{ii}h_{jj}\Bigr)\notag\\[-0.25em]
&\quad+[\partial\pi^2]\Bigl(
-h_{00}^2-4c_s^2h_{0i}h_{0i}+2c_s^2h_{ij}h_{ij}
+2c_s^2h_{00}h_{ii}+(f_3+c_s^2)h_{ii}h_{jj}\Bigr)\notag\\
&\quad+\dot\pi^2\Bigl(
3h_{00}^2-4c_s^2h_{0i}h_{0i}+2c_s^2h_{ij}h_{ij}
-2c_s^2h_{00}h_{ii}+(f_3+c_s^2)h_{ii}h_{jj}\Bigr)\notag\\
&\quad-8[\partial\pi]\Bigl(
(1+c_s^2)h_{00}-(f_3+2c_s^2)h_{ii}\Bigr)h_{0j}\dot\pi^j
+8(1-c_s^2)(h_{0i}\dot\pi^i)^2\notag\\[-0.25em]
&\quad+8(h_{00}-c_s^2h_{ii})h_{0j}\dot\pi^k\partial_k\pi^j
+4(h_{00}-c_s^2h_{kk})h_{ij}\dot\pi^i\dot\pi^j
\biggr\},\\[.7 em]
\mathcal L_{h^3\pi}={}&\frac{\kappa^3}{48\lambda}\biggl\{
[\partial\pi]\Bigl(
-3h_{00}^3+3c_s^2h_{00}^2h_{ii}
-3(f_3+c_s^2)h_{00}h_{ii}h_{jj}\notag\\
&\hspace{3em}-(3f_3+f_4+c_s^2)h_{ii}h_{jj}h_{kk}
+12\bigl((f_3+c_s^2)h_{ii}-c_s^2h_{00}\bigr)h_{0j}h_{0j}
-6\bigl(c_s^2h_{00}+(f_3+c_s^2)h_{ii}\bigr)h_{jk}h_{jk}
\notag\\
&\hspace{3em}
+24c_s^2h_{0i}h_{ij}h_{0j}
-8c_s^2h_{ij}h_{jk}h_{ki}\Bigr)\notag\\
&\quad+6h_{0i}\dot\pi^i\Bigl(
3h_{00}^2-4c_s^2h_{0j}h_{0j}+2c_s^2h_{jk}h_{jk}
-2c_s^2h_{00}h_{jj}+(f_3+c_s^2)h_{jj}h_{kk}\Bigr)
\biggr\}, \\[.7 em]
\mathcal L_{h^4}={}&-\frac{\kappa^2}{2}\biggl\{
\frac18\left(h^2-2h_{\lambda\tau}h^{\lambda\tau}\right)
 \eta^{\mu\nu}\eta^{\alpha\beta}\eta^{\rho\sigma}
-\frac{h}{2}\Bigl(
h^{\mu\nu}\eta^{\alpha\beta}\eta^{\rho\sigma}
+\eta^{\mu\nu}h^{\alpha\beta}\eta^{\rho\sigma}
+\eta^{\mu\nu}\eta^{\alpha\beta}h^{\rho\sigma}
\Bigr)
\notag\\
&\hskip 1.15 cm 
+(h^2)^{\mu\nu}\eta^{\alpha\beta}\eta^{\rho\sigma}
+\eta^{\mu\nu}(h^2)^{\alpha\beta}\eta^{\rho\sigma}
+\eta^{\mu\nu}\eta^{\alpha\beta}(h^2)^{\rho\sigma}
+h^{\mu\nu}h^{\alpha\beta}\eta^{\rho\sigma}
+h^{\mu\nu}\eta^{\alpha\beta}h^{\rho\sigma}
+\eta^{\mu\nu}h^{\alpha\beta}h^{\rho\sigma}
\biggr\}
\notag\\
&\qquad\times\biggl(
\partial_\mu h_{\alpha\beta}\partial_\nu h_{\rho\sigma}
-\partial_\mu h_{\alpha\rho}\partial_\nu h_{\beta\sigma}
+2\partial_\mu h_{\beta\rho}\partial_\alpha h_{\nu\sigma}
-2\partial_\mu h_{\nu\alpha}\partial_\beta h_{\rho\sigma}
\biggr),
\\[.7 em]
\mathcal L_{h^2\chi^2}={}&\frac{\kappa^2}{2}\biggl\{\left[(h^2)^{\mu\nu}-\frac{h}{2}h^{\mu\nu}\right]\partial_\mu\chi\partial_\nu\chi+\frac18\left(h^2-2h_{\rho\sigma}h^{\rho\sigma}\right)\left(\partial_\mu\chi\partial^\mu\chi-m^2\chi^2\right)\biggr\}
\end{align}
\end{subequations}
\endgroup

Though these interaction terms are what one gets from expanding in $D=4$, some can be written more simply as the part that contributes to the Feynman rule, plus a total derivative, which is a boundary term in the action that does not contribute to the bulk Feynman rule. For example, the cubic phonon interaction can be rewritten as
\begin{align}
    \mathcal L_{\pi^3}^{(D=4)}={}&\lambda
    \left\{
    \frac{c_s^2}{2}
    [\partial\pi][\partial\pi^2]
    -
    \frac{3c_s^2+f_3}{6}
    [\partial\pi]^3
    +
    \frac{1+c_s^2}{2}
    [\partial\pi]\dot{\pi}^{\,2}
    -
    \dot{\pi}^{\,i}\partial_i\pi^j\dot{\pi}^{\,j}
    \right\} 
    + \lambda\; \partial_i
    \left(
    \frac{1}{6}
    \tilde\epsilon^{ijk}\tilde\epsilon_{abc}\,
    \pi^a\partial_j\pi^b\partial_k\pi^c
    \right).
\end{align}
The first grouping $\lambda \{\cdots\}$ is actually the cubic phonon interaction Lagrangian in $D=3$. One can show more generally that $\mathcal L_{\pi^n}$ expanded in dimension $D=n$ differs from that obtained in dimension $D > n$ only by a total derivative term. Thus, the bulk $V_{\pi^n}$ Feynman rule is the same in all dimensions $D \geq n$. 


\section{Four-point Feynman rules and generalized Ward identities}
\label{app:four-point-ward-identities}

\begin{figure}[p]
\centering
\ManuscriptFigurePDF[scale=1]{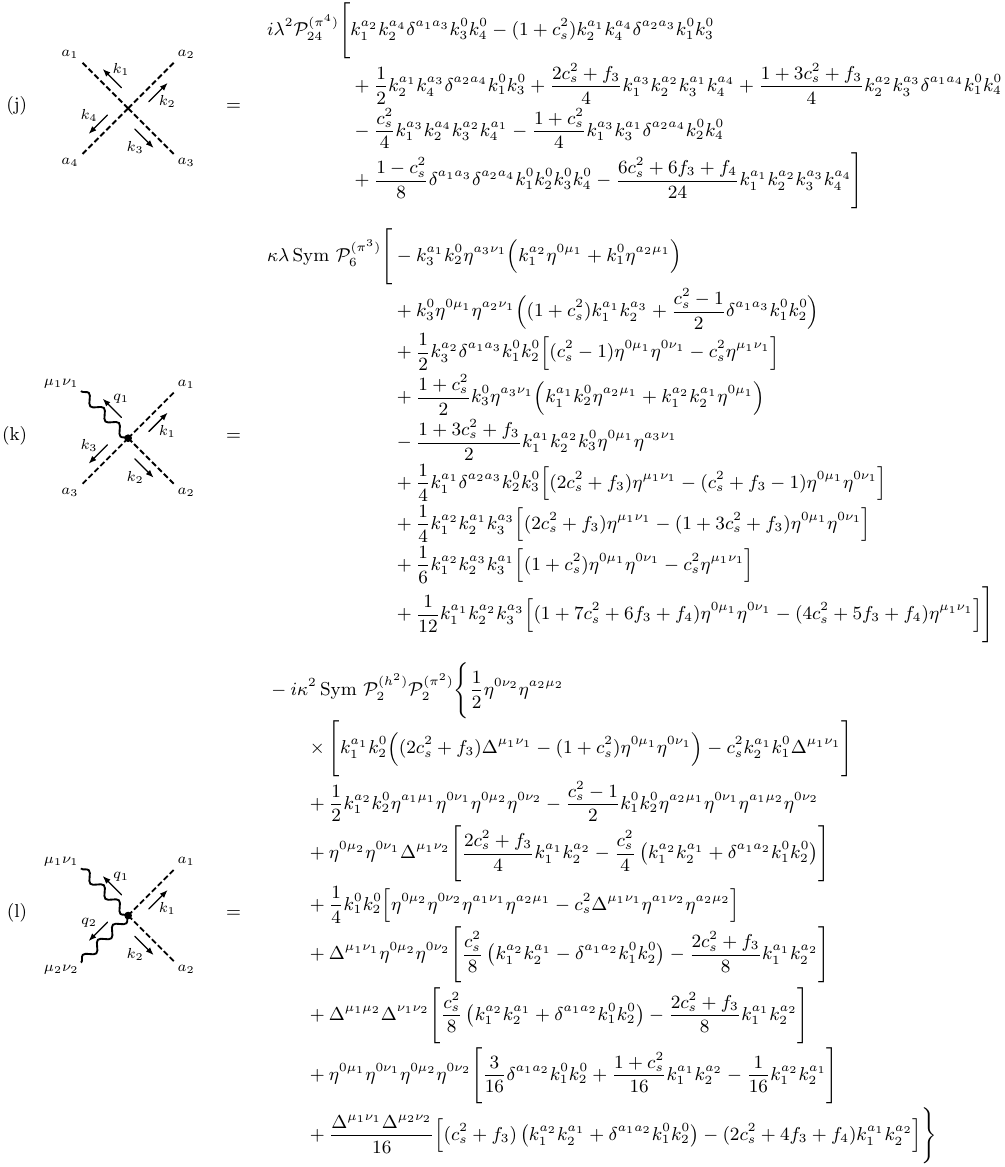}
\caption{Four-point vertices:
$\pi^4$, $h\pi^3$, $h^2\pi^2$, $h^3\pi$, $h^4$, and $h^2 \chi^2$. Filled circles $\bullet$ denote vertices with both phonons and gravitons, while vertices that are pure-phonon, pure-gravity, or gravity-probe interactions do not have an explicit vertex dot.}
\label{fig:four-point-vertices}
\end{figure}

Here we collect the four-point Feynman vertices, which first enter two-body calculations at two loops. These appear in our check of the four-point on-shell Ward identities, which are very useful for confirming that the vertices are correctly derived.   They are also given in a plain-text Mathematica ancillary file, 
\texttt{FeynmanRules.m}, together with all other propagators and interaction vertices through four points for this EFT.  \autoref{fig:four-point-vertices} reports the four-point Feynman rules, following the operator and Wilson coefficient conventions of \app{app:conventions}. We also define the spatial projector to write the vertices compactly,
\begin{equation}
\Delta^{\mu\nu}\equiv \sum_{i=1}^{D-1}\eta^{i\mu}\eta^{i\nu}.
\label{eq:spatial-projector-four-point}
\end{equation}

\begin{figure}[p]
\centering
\ManuscriptFigurePDF[scale=1]{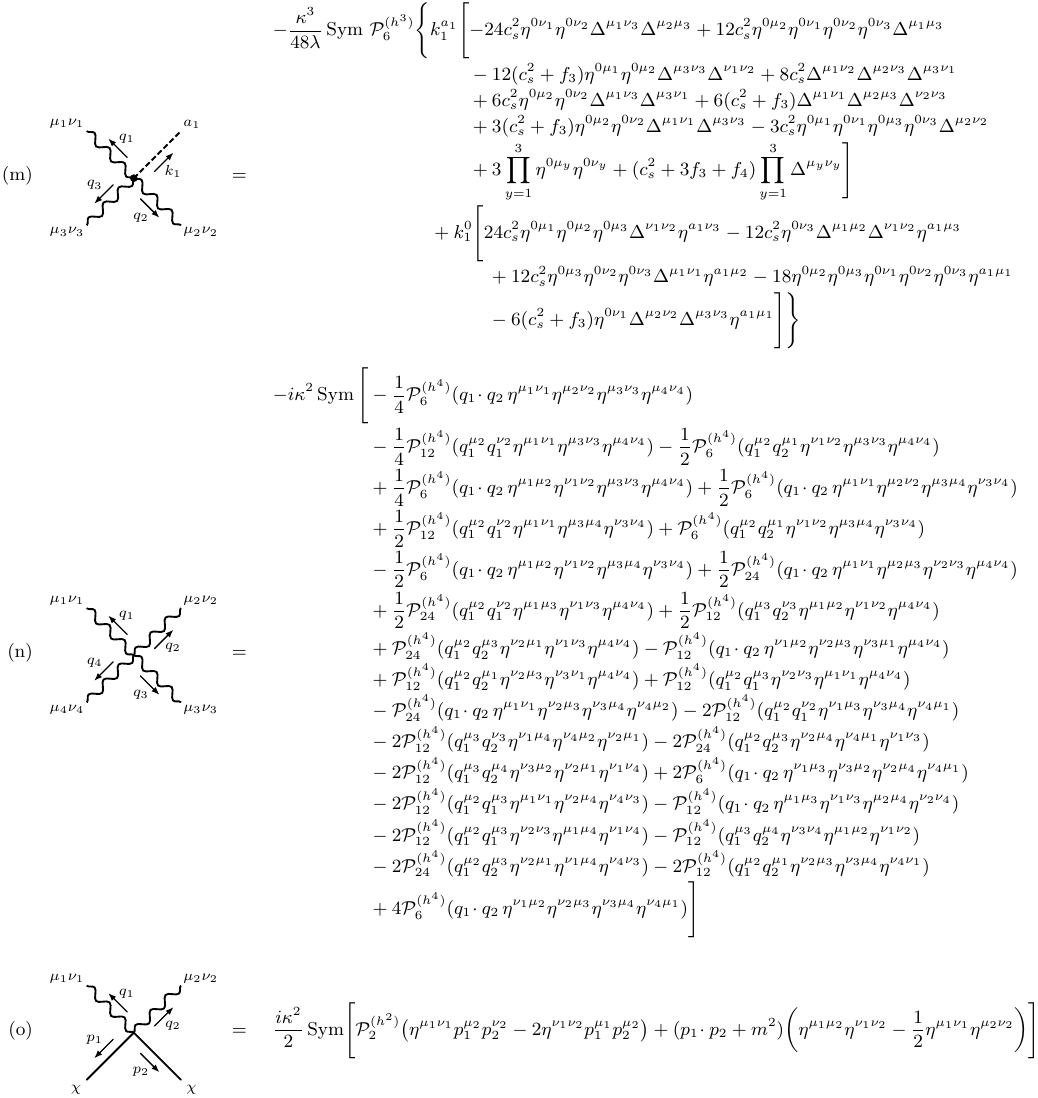}
\ContinuedFigureCaption{Four-point vertices (continued).}
\end{figure}
\clearpage

\begin{figure}[!t]
\centering
\ManuscriptFigurePDF[scale=1]{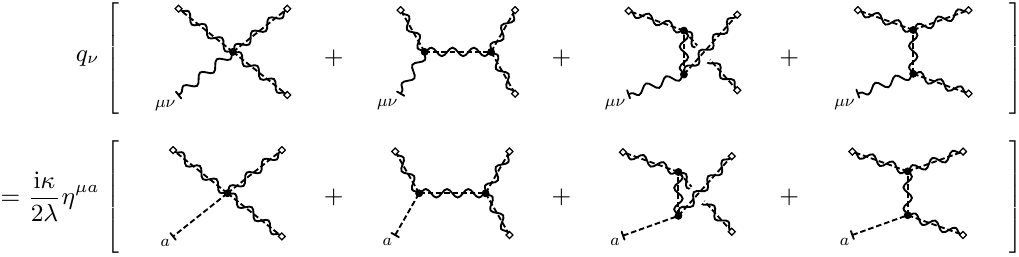}
\vspace{1.0em}
  \caption{Four-point on-shell Ward identity in the
  full graviton-phonon multiplet basis. All spectators marked by diamonds $\diamond$ are on shell and the root momentum
  $q$ remains open. The upper and lower rows carry an open graviton and
  phonon root, respectively.  Each row shows the quartic contact followed by the rooted $s=1|23$, $t=2|13$, and $u=3|12$ exchange
  channels.}
\label{fig:4-pt-multiplet-ward}
\end{figure}

\begin{figure}[p]
\centering
\ManuscriptFigurePDF[scale=1]{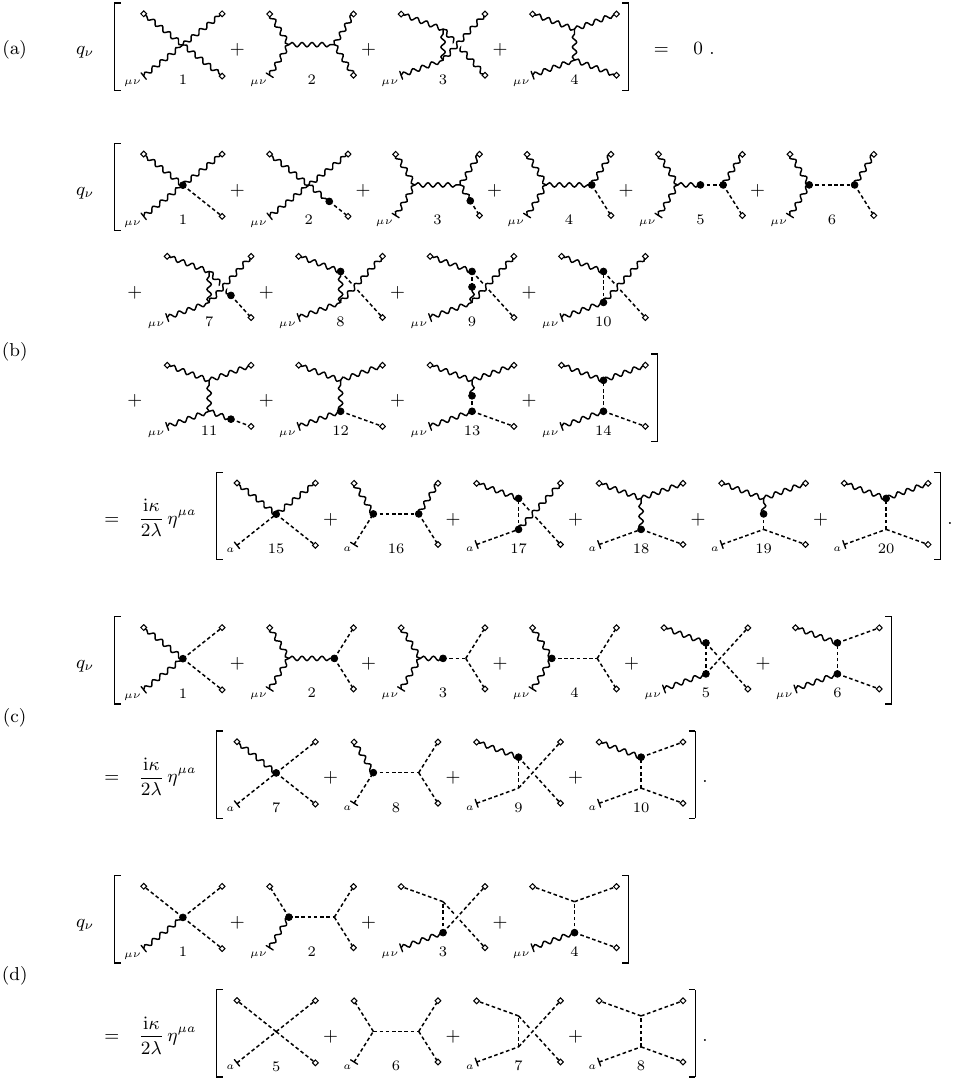}
  \vspace{1.8em}
  \caption{Four-point on-shell Ward identities in
  the non-interacting basis at leading order in $\kappa/\lambda$.  Rows (a)--(d) are the Einstein--Hilbert $h^4$,
  $h^3\pi$, $h^2\pi^2$, and $h\pi^3$ identities.  Wavy and dashed
  lines are pure gravitons and phonons, filled circles $\bullet$ denote fluid--gravity interactions, diamonds $\diamond$ mark the three on-shell
  spectators, and every barred open root carries momentum $q$.   Across every contact and exchange graph, the barred
  open root is fixed at lower left, while spectators $1,2,3$ occupy upper
  left, upper right, and lower right, respectively.  The horizontal, crossed,
  and vertical exchange connectivities are the $s=1|23$, $t=2|13$, and
  $u=3|12$ rooted channels, respectively.}
\label{fig:4-pt-diagonal-ward}
\end{figure}

The four-point on-shell Ward identity at the level of the 13-component graviton-phonon multiplet is shown diagrammatically in \autoref{fig:4-pt-multiplet-ward}.
The current includes a four-point contact interaction, along with the three exchange $s$, $t$, and $u$ channels between two three-point interactions.  At leading order in the weak-mixing perturbative expansion, the resulting four-point on-shell Ward identities in the non-interacting basis are listed in \autoref{fig:4-pt-diagonal-ward}. The explicit checking of these non-interacting basis on-shell Ward identities at tree-level four-points is given in the ancillary file \texttt{FourPointWardCheck.m}. In this ancillary file, the open index $\mu$ is contracted with the unconstrained auxiliary four-vector $\xi_\mu$, so each of the 42 diagrams after contraction is just a Lorentz scalar, which is reported explicitly.  Let the open root have outgoing momentum $q$ and the upper-left, upper-right, and lower-right spectators have the respective outgoing momenta and physical polarization pairs $(k_1,e_1)$, $(k_2,e_2)$, and $(k_3,e_3)$.  Then, imposing momentum conservation $q+k_1+k_2 +k_3=0$ and the on-shell conditions of the physical states outlined in \autoref{sec:physical_state_conditions}, the ancillary file shows that each generalized Ward identity in \autoref{fig:4-pt-diagonal-ward} is satisfied at leading order in $\kappa/\lambda$.

Notably, at leading order in weak mixing, the four-point pure gravity on-shell Ward identity \autoref{fig:4-pt-diagonal-ward}(a) is unchanged at tree level, reflecting the general conclusion in the text about the preservation of this identity at $n$-points. The modifications to the on-shell Ward identities for mixed vertices can be identified by following the procedure outlined in \autoref{sec:ScatteringAmplitudes} for writing down a diagram with fixed external legs in a specified channel with minimal mixing (lowest order in $\kappa / \lambda$). This procedure is equivalent to successively ``pushing'' the external phonons further into a diagram in a specified channel, while preserving the same order in the weak mixing parameter. There are very few diagrams in each channel that must be written down by the selection rule shown in \autoref{fig:n-Point_TT_ConvertVanish}, since external TT gravitons cannot mix into a phonon in the fluid rest frame.  Thus, in the non-interacting basis expansion with TT gravitons, explicit mixed or conversion legs are restricted to phonon LSZ spectators or internal exchange propagators.

\clearpage

\bibliography{references}


\end{document}